\documentclass[prd,twocolumn,draft]{revtex4}
\usepackage{bm}
\usepackage{hyperref}
\usepackage{latexsym}
\usepackage{mathrsfs}
\usepackage{graphicx}
\usepackage{graphics}
\usepackage{txfonts}
\usepackage{fancybox}
\usepackage{here}
\usepackage{helvet}
\usepackage{color}
\usepackage{fancyhdr}
\input epsf.sty
{

\font\titlefont=cmssbx18 

\font\authorfont=cmr12
\font\abstractfont=cmr10

\font\sectionfont=cmssbx9
\font\subsectionfont=cmssbx8
\font\subsubsectionfont=cmssbx8

\font\figtextfont=cmss9
\font\tabletitlefont=cmssbx10 
\font\tabletextfont=cmss9
\font\reftitlefont=cmssbx9
\font\reftextfont=cmr10

\begin{document}

\title{\titlefont On a New 4-Vector Cosmological Field Theory}

\author{\authorfont\textbf{ G. G. Nyambuya}}
\email{gadzirai@gmail.com, fskggn@puk.ac.za}

\affiliation{%
\authorfont  North-West University -Potchefstroom Campus, School \textit{of} Physics - Unit \textit{for} Space Research, Private Bag X6001, Potchefstroom 2531, Republic \textit{of} South Africa. \\} 

\date{\today}

\begin{abstract}
\begin{center}

\end{center}
\linethickness{2pt}
\line(1,0){400}\\
\\
\textbf{\sectionfont Abstract.} {\abstractfont The original Dirac Equation  is modified in the simplest imaginable and most trivial manner to include a universal \textit{$4$-Vector Cosmological Field} term in the space and time dimensions. This cosmological  field leads to a modified Dirac Equation  capable of explaining why the Universe appears to be made up chiefly of matter. It is seen that this \textit{$4$-Vector Cosmological Field} is actually a particle field and this particle field can possibly be identified with the darkmatter and darkenergy field. Further, this \textit{$4$-Vector Cosmological Field} is seen to give spacetime the desired quantum mechanical properties of randomness. Furthermore, it is seen that in the emergent Universe, the position coordinates of a particle in space -- contrary to the widely accepted belief that the position of a particle in space has no physical significance, we see that that opposite is true -- namely that the position of a particle has physical significance. We further note that the \textit{$4$-Vector Cosmological Field} modification to the Dirac Equation leads us to a vacuum model redolent but differrent from that of Quantum Electrodynamics (QED). This new vacuum model is without virtual particles but darkparticles. We dare to make the suggestion that these darkparticles may possibly explain the current mystery of what really is darkmatter and darkenergy.}
\line(1,0){400}

\textsl{\begin{center}
``\textbf{We have found a strange footprint on the shores of the unknown.\\ We have devised profound theories, one after another, to account for its origins.}''
\end{center}}

\begin{flushright}
-- \textbf{Sir Arthur S. Eddington} (1882 - 1944)
\end{flushright}
\end{abstract}

\maketitle

\section{\sectionfont Introduction}

Each electronically charged elementary particle has a counterpart with the opposite electronic charge which is known as its \textit{antiparticle} (antiparticles are also referred to as \textit{antimatter}) and just like normal particles, antiparticles do combine, forming atoms of antimatter which some call antiatoms --  albeit, unlike atoms, these do not live long. Paul Dirac's brilliant theory proposed in 1928 predicted the existence of antimatter (Dirac $1928a,b$). It [Dirac's Theory] is one of the most successful Theories \textit{of} Physics. This theory, suggested that the Laws \textit{of} Nature are exactly the same for matter and antimatter; so given this symmetry, the Universe must contain matter and antimatter in equal propositions everywhere and everytime -- that is, across all of spacetime.  Unfortunately (or maybe fortunately -- as will be argued soon), when we look into our immediate vicinity, we see that this is not the case -- our terrestrial habitate seems to be dominated exclusively by matter; so the question  \textsl{``Why is our measurable Universe made up chiefly of matter  with no significant quantities of antimatter?''}  has always been hanging in-limbo since Dirac's theory was set forth -- for a good review of the origins and possible solutions to the problem of matter-antimatter asymmetry see e.g. Due \& Kusenko ($2004$).

While we may wonder why the Universe is formed this way, \textit{viz} matter-antimatter imbalance, we must be very thankful that the Universe is formed this way, because if it [Universe] did really have equal proportions of matter and antimatter uniformly distributed throughout all of space and time, you the reader would not be reading this because the Universe would be nothing but a hot-bath of radiation because matter and antimatter would annihilate to form radiation. Despite its great success. The search for an answer to this great cosmic mystery -- \textit{why we are so lucky to have a Universe chiefly made-up of matter} -- is the main theme of the present reading and it is important to mention that our adventures to seeking an answer to this great cosmic mystery will take us to other areas of physical enquiry and new discoveries. Though it shall prove difficult, we shall try to not veer too much off the main road but keep as much as we can to what we want to achieve here.

It is worthwhile to mention here that the first and probably current-best  attempt at an answer to this question is that by the Russian Physicist, father of the hydrogen bomb and $1975$ Nobel Peace Prize winner, Andrei Sakharov ($1924-1987$). The attempt by Andrei Sakharov ($1967$) is the widely accepted explanation as to why there exist this matter-antimatter asymmetry -- we offer an asymptotically different solution! He [Andrei Sakharov] argued, that to create an imbalance between matter and antimatter from an initial condition of balance, certain conditions must be met and these conditions have come to the called the Sakharov conditions and $\textrm{CP}$\textit{-violation} is one of the conditions. $\textrm{CP}$\textit{-violation} is a violation of the symmetry where the Laws \textit{of} Nature are expected to act the same when we simultaneously interchange the electronic charge ($\textrm{C}$-\textit{symmetry} known as charge conjugation symmetry) of a particle and invert the space coordinates $\textrm{P}$-\textit{symmetry} (known as parity symmetry).  

Given the need for $\rm{CP}$\textit{-asymmetric} equations in physics, much to the dismay of the physicist, the Fundamental Equations  \textit{of} Physics, in their bare form, do not exhibit $\textrm{CP}$\textit{-violation} (or $\rm{P}$\textit{-violation}) and this -- sadly and against the desiderata, has to be inserted by hand into the equations. For example, in the Standard Model \textit{of} Particle Physics, the Cabibbo-Kobayashi-Maskawa matrix (Cabibbo $1963$; Kobayashi \& Maskawa $1973$) is employed and a complex phase factor is artificially injected into this matrix to bring about $\textrm{CP}$\textit{-violation} inorder to explain the observed $\rm{CP}$\textit{-violation} in the Kaon system. $\textrm{CP}$\textit{-violation} is observed in Kaons -- for a good read on the history of the discovery of this see e.g. Lacoste-Julien ($2003$). $\textrm{CP}$\textit{-violation} has been observed in B-meson aswell (see e.g. Aubert \textit{et. al.} $2001$). Given Sakhorov's thesis, this $\rm{CP}$\textit{-violation} (in the B-meson and Kaon system) is thought of as holding the key to unlocking the mystery of matter-antimatter asymmetry albeit some researchers (e.g. Rodgers $2001$; Sinha $2009$)  feel it [the observed $\textrm{CP}$\textit{-violation} in the B-meson and Kaon system] is not enough to explain the observed matter-antimatter asymmetry.

In this reading, we make a modification to the Dirac Equation by the addition of a \textit{4-Vector Cosmological Field} and from this, we demonstrate that this modification leads to an equation that clearly points to the fact that a stable Universe can only have one form of matter; either it is filled with matter or antimatter. Further, the emergent Universe from our equations is such that if antimatter is to exist in a Universe of matter, then, it will not be stable thus it has to decay. 

The possibility of the existence of a cosmological  field has sound justification and it is inferred from the cosmological observations such as the apparent accelerated expansion of the Universe and the indication from the rotation curves of galaxies that there must exist a form of unseen matter or energy. This unseen matter or energy is popularity known as the Darkmatter/Darkenergy respectively. If our equations are correct, then, this dark matter/energy can be identified with the cosmological  field (which is better here referred to as a cosmic fluid) and it is seen that this cosmic fluid is a fluid composed of pure-point particles, that is -- particles with no breath, height nor length thus are susceptible to be permeable and all-pervading. These particles making up the cosmic fluid travel at the speed of light. It is also seen that this cosmic fluid acts as a barrier forbidding positive energy particles from falling into negative energy states. Actually a new vacuum model is set forth.

I would like to say here, that this reading is written with the full knowledge of the results of our other readings (Nyambuya $2008a,b$) where we have presented, what we believe is a viable  curved spacetime version of the Dirac Equation. Our focus here is the original Dirac Equation which is a subset of the proposed Curved Spacetime Dirac Equations (Nyambuya $2008a,b$).  This has been done for the sole reason that we would like to show from the well and universally accepted Dirac Equation that  matter and antimatter asymmetry can be explained on the basis of an all-pervading and permeating cosmic fluid (cosmological  field). Having shown that  matter-antimatter asymmetry can be explained on this basis of an all-pervading and permeating cosmic fluid, we hope in the very near future to use this to further modify the equations presented in (Nyambuya $2008a,b$). 

\section{\sectionfont The Dirac Equation and its Symmetries}

Without being to assuming, let us, for instructive purposes browse through  the thesis leading to the Dirac Equation. Suppose we have a particle of rest-mass $m_{0}$ and momentum $p$ and energy $E$, Albert Einstein, from his $1905$ Special Theory \textit{of} Relativity (STR) -- derived the basic equation:

\begin{equation}
E^{2}=p^{2}c^{2}+m_{0}^{2}c^{4}.\label{Emc2}
\end{equation}

This equation formed the basis of the Klein-Gordon Theory upon which the Dirac
Theory is founded.  Using the already established canonical quantisation procedures, $\vec{\textbf{p}}_{\mu}\longrightarrow i\hbar\partial_{\mu}$, Klein and Gordon proposed the equation:

\begin{equation}
\square\Psi=\left(\frac{m_{0}c}{\hbar}\right)^{2}\Psi,\label{Klein-Gordon 1}
\end{equation}

where $\hbar$, $c$ are the Planck constant, the speed of light respectively, an $\Psi$ a scalar wavefunction while $ \square=\partial^{2}/c^{2}\partial t^{2}-\nabla^{2}$. This equation describes a spin-$0$ quantum mechanical scalar particle and allows for  negative probabilities which from a physical standpoint are meaningless, and for this reason, Dirac was not satisfied with the Klein-Gordon Theory. He noted that the Klein-Gordon equation is second order differential equation and his suspicion was that the origin of the negative probability solutions may have something to do with this very fact. 

He sought an equation linear in both the time and spatial derivatives that would upon ``squaring" reproduce the Klein-Gordon equation. The equation he found was:

\begin{equation}
\left[i\hbar\gamma^{\mu}\partial_{\mu}-m_{0}c\right]\psi=0,\label{Dirac}
\end{equation}

where: 

\begin{equation}
\begin{array}{c c}
\gamma^{0}=
\left(\begin{array}{c c}
\textbf{I} & \boldsymbol{0}\\
\boldsymbol{0} & -\textbf{I} \\
\end{array}\right)
,\,\,\,\,
\gamma^{i}=
\left(\begin{array}{c c}
\textbf{0} & \boldsymbol{\sigma}^{i}\\
-\boldsymbol{\sigma}^{i} & \textbf{0} \\
\end{array}\right)
\end{array},
\end{equation}

are the  $4\times4$ Dirac gamma matrices ($\textbf{I}$ and $\textbf{0}$ are the 2$\times$2 identity and null matrices respectively) and $\psi$ is the four component Dirac wave-function, namely:

\begin{equation}
\psi=\left(\begin{array}{c}
\psi_{0}\\
\psi_{1}\\
\psi_{2}\\
\psi_{3}\end{array}\right)=\left(\begin{array}{c}
\Phi\\
\chi\\
\end{array}\right),\,\rm{where:}\,\begin{array}{c}
\Phi=\left(\begin{array}{c}
\psi_{0}\\
\psi_{1}\\
\end{array}\right)\\
\\
\chi=\left(\begin{array}{c}
\psi_{2}\\
\psi_{3}\end{array}\right)
\end{array} \label{4spinor}.
\end{equation}

Throughout this reading -- unless otherwise specified, the Greek indices will be understood to mean $\mu,\nu, ... = 0,1,2\,\rm{or}\,3$ and the lower case English alphabet $i,j, ... = 1,2\, \rm{or}\, 3$. In (\ref{4spinor}), the first representation of $\psi$ is known as the four component representation and the second in which this spinor is written in-terms of $\Phi$ and $\chi$, is the bi-spinor representation.

The Dirac Equation is perfectly symmetric equation -- \textit{viz}, it obeys the following symmetries:

\begin{enumerate}
\item Obeys C-\textit{symmetry}. This is symmetry under the interchange of the electronic charge of the particle.

\item Obeys $\rm{T}$-\textit{symmetry}. This is symmetry under the interchange of the hand of time in the Dirac Equation, that is $t\longmapsto -t$.

\item Obeys P-\textit{symmetry}. This is symmetry under the interchange of the space coordinates in the Dirac Equation, that is $x^{\mu}\longmapsto -x^{\mu}$.

\item Obeys $\rm{CT}$-\textit{symmetry}. This is symmetry under  both $\rm{C}$ and $\rm{T}$.

\item Obeys $\rm{CP}$-\textit{symmetry}. This is symmetric under  both $\rm{C}$ and $\rm{P}$.

\item Obeys $\rm{PT}$-\textit{symmetry}. This is symmetry under  both $\rm{P}$ and $\rm{T}$.

\item Obeys $\rm{CPT}$-\textit{symmetry}. This is symmetry under  all the operations $\rm{C}$, $\rm{P}$ and $\rm{T}$.

\item Obeys Lorentz invariance. This is symmetry under the change of the inertial frame of reference.

\end{enumerate}

We shall not demonstrate these symmetries but direct the reader to Nyambuya ($2008a,b$) or any good book of Quantum Mechanics (QM) that deals with Dirac's Equation.

Now if we are to add a cosmological  field, $\pm\Lambda_{0}$ (this constant has the dimensions of inverse length and we shall assume it to be real $\Lambda_{0}>0$) to the energy ($E$) of a particle, and the cosmological  field will be assumed to be an all-pervading and permeating form of energy that fills all of space at everytime -- we will have to make the transformation $E\longrightarrow \mathcal{E}\pm\Lambda_{0} \hbar c$ because this energy will add to the existing energy $E$. This modification, $E\longrightarrow \mathcal{E}\pm\Lambda_{0} \hbar c$,  leads to equation (\ref{Emc2}) transforming to:

\begin{equation}
\mathcal{E}=\pm\Lambda_{0} \hbar c\pm\sqrt{p^{2}c^{2}+m_{0}^{2}c^{4}},\label{Emc3}
\end{equation}

hence thus the modification we seek to make to the Dirac Equation must, as the Dirac Equation upon ``squaring'', lead us to this equation. For most of the times, we shall consider the case $+\Lambda_{0}$ without considering the case  $-\Lambda_{0}$ as considering one case, is as good as the other has been considered and this is because of the symmetric nature of the equations for these two cases. 

If as in the Dirac formulation, a particle of negative energy is the antiparticle, then according to the (\ref{Emc3}), it would mean that antiparticles and particles must -- unlike the Dirac theory, have unequal energies hence unequal masses since $\mathcal{E}_{+}\neq |\mathcal{E}_{-}|$ where $\mathcal{E}_{+}$ and $\mathcal{E}_{-}$ are the positive and negative energy solutions of (\ref{Emc3}) respectively. This is clearly contrary to observations. What does this mean for the theory we wish to set forth? Is it stillborn? How would one explain the observed equality in mass of particles and antiparticles? We shall address this question in the next section.

For the sake of completeness, we shall veer a little on to the side of history. The idea of a cosmological field originally dates back to Einstein, albeit Einstein's cosmological  field is a scaler whereas the present is a four vector. After  discovering his [Einstein] now famous Law \textit{of} Gravitation (initially with $\Lambda=0$), namely:

\begin{equation}
R_{\mu\nu}-\frac{1}{2}Rg_{\mu\nu}=\kappa T_{\mu\nu}+\Lambda g_{\mu\nu},\label{Einstein's Field eqn}\end{equation}

where: $R_{\mu\nu}$, is the Riemann curvature tensor and $R$ is the contracted Riemann curvature tensor and: 

\begin{equation}
T_{\mu\nu}=\varrho v_{\mu}v_{\nu}+pg_{\mu\nu},\label{stress-tensor}
\end{equation}

is the stress and energy tensor where $\varrho$ is the density of matter, $p$ is the pressure and $v_{\mu}$ the four velocity, $\kappa=8\pi G/c^{4}$ is the Einstein constant of Gravitation with $G$ being Newton's Universal Constant of Gravitation, Einstein added $\Lambda$ controversial scalar cosmological  field term (with SI units of $\textrm{m}^{-2}$). He did this so as to ``stop'' the Universe from expanding (Einstein $1917$) and he was motivated to do so because of the strong influence from the astronomical wisdom of his day that the Universe appeared to be static and thus was assumed to be so. Besides, the cosmological  field fullfiled Mach's Principle (Mach $1893$), a principle that had inspired Einstein to search for the GTR and thus he thought that the GTR will have this naturally embedded in it. Mach's principle forbids the existence of a truly empty space and at the sametime supposes that the inertia of an object is due to the induction effect(s) of the totality of all-matter in the Universe. 

We introduce here the cosmological  field for a reason asymptotically different from that of Einstein, namely that we wish to explain the asymmetry between matter and antimatter and not to ``stop'' the Universe from expanding. As will be seen -- in the presence of an ambient electromagnetic field, our introduction of the cosmological  field will induce an asymmetry in the time dimension and this asymmetry naturally leads to a modified Dirac Equation that violates $\rm{C}$-\textit{symmetry} and also the combined charge and time symmetry, i.e $\rm{CT}$-\textit{symmetry}.

\section{\sectionfont Connection Between Electronic Charge \& Rest-mass }

As already pointed out, if as in the Dirac formulation, where the particle with positive energy $\mathcal{E}_{+}$ is considered to be the particle; and the one with negative energy $\mathcal{E}_{-}$ is considered to be the antiparticle -- then, according to the present ideas, it would mean that antiparticles and particles must have unequal energies since $\mathcal{E}_{+}\neq|\mathcal{E}_{-}|$  as is, in the case of the Dirac Theory. According to the Einstein mass-energy  equivalence ($\mathcal{E}=mc^{2}$), the masses of particles and antiparticles will not be equal too. This is clearly contrary to observations as particles and antiparticles have been observed to have equal masses. 

In the reading Nyambuya ($2008b$), an attempt at this question has been made, where the electronic charge ($Q$) of a fundamental particle has been related to its rest-mass, that is $m_{0}\propto Q\Longrightarrow m_{0}c=\epsilon Q$ where $Q$ is the electronic charge of the fundamental particle and $\epsilon$ some numerical constant. This means if the rest-mass of a particle is $+m_{0}$, its antiparticle's rest-mass will be $-m_{0}$, and these will have the same energy solution $\mathcal{E}$ since the particle's energy will be given by $\mathcal{E}=\Lambda_{0} \hbar c+\sqrt{p^{2}c^{2}+(+m_{0})^{2}c^{4}}$ and its antiparticle's energy will be $\mathcal{E}=\Lambda_{0} \hbar c+\sqrt{p^{2}c^{2}+(-m_{0})^{2}c^{4}}$ which are exactly the same hence they will have the same mass. 

This conclusion, that $m_{0}c=\epsilon Q$, was reached after consideration of the derived curved spacetime Dirac Equation there-in Nyambuya ($2008b$), where it was required that this derived curved spacetime Dirac Equation, remain invariant under the reversal of the rest-mass, that is $m_{0}\longmapsto-m_{0}$. The Dirac Equation is symmetric under a reversal of the rest-mass of the particle but this symmetry is of little significance in the Dirac Theory. In Nyambuya ($2008b$), we found out that the transformation $m_{0}\longmapsto-m_{0}$ entails the reversal of the particle's electromagnetic field and from this we argued that the rest-mass must have an intimate connection with the electronic charge of the particle. We thus direct the reader to this reading Nyambuya ($2008b$) for the full argument on this.

Taking as given that, $m_{0}c=\epsilon Q$, it means we can write equation (\ref{Emc3}) as:

\begin{equation}
\mathcal{E}=\pm\Lambda_{0} \hbar c\pm\sqrt{p^{2}c^{2}+\epsilon^{2}Q^{2}c^{2}}.\label{Emc4}
\end{equation}

It is clear from this equation that under this relationship -- $m_{0}c=\epsilon Q$, a particle and its antiparticle will have the same energy ($\mathcal{E}$) and hence the same mass ($m=\mathcal{E}/c^{2}$). 

One may want to argue that if the electronic charge of a fundamental particle is related to it's rest mass in the manner suggested here, then, the magnitude of the Electron and the Proton's electronic charge when at ``rest'' must be equal. But we know from the uncertainty principle of QM ($\Delta p\cdot\Delta x\sim \hbar$) that for a particle of finite dimensions ($\Delta x$) like the Electron and Proton, the concept of rest is without meaning since bringing these particles into a ``true state of rest'' ($\Delta p=0$) would mean the dimensions these particles would have t be infinite ($\Delta x=\hbar/\Delta p=\hbar/0=\infty$). In this way, QM mechanics informs us that a particle of finite dimensions  can not be at rest hence thus the thinking that an Electron and a Proton can be brought to rest is null and void. It may appear to us to be at rest -- that is, the Electron and Proton may appear to us to be stationed at the same position, but if we had a way to magnify this to a magnification of our liking, then according to QM, this particle will be seen to be dotting up and down randomly meaning to it has momentum hence thus the whole thinking of an Electron and/or Proton at rest is obsolete.

Neglecting the negative energy solution, equation (\ref{Emc2}) may be written $E=|m_{0}c^{2}|(1+p^{2}/|m_{0}c^{2}|)^{1/2}$ and in the case where the momentum of the particle is small (as in the above case of the Electron and the Proton), that is $p^{2}\lll m_{0}^{2}c^{2}$, to first order approximation this reduces to $E\simeq p^{2}/2|m_{0}|+|m_{0}|c^{2}$, thus for the Electron and Proton, will have $\Delta m=E_{p}-E_{e}=(p^{2}_{p}-p^{2}_{e})/2|m_{0}|$ where $E_{p}, E_{e},p_{p}$ and $p_{e}$ are the Electron and Proton's energy and momentum respectively. Hence thus, the difference in the Electron and the Proton's mass is a measure of the difference in the square of their momentum.

\section{\sectionfont Partial Cosmological Dirac Equation}

There exists two avenues on which to arrive at equation (\ref{Emc4}). We prefare the form (\ref{Emc4}) than (\ref{Emc3}) because (\ref{Emc4}) makes it clear where the electronic charge ($Q$) of the particle fits in explicitly hence it will be easy to investigate its symmetries under the intercharge of the particle's electronic charge. Thus from here-on, we shall understand that:  $m_{0}c=\epsilon Q$, hence thus a reversal of the particle's electronic charge is a reversal of the particle's rest-mass. 

As will be seen, we shall have to investigate the symmetries of the equations that we shall derive here and in the cases where charge reversal symmetry is concerned, we shall have to deal with the electromagnetic properties of the particle by reversing these properties. In Nyambuya (2008a), we derived three Dirac Equations for curved spacetime and amongst these is the equation:

\begin{equation}
\left[i\hbar\Gamma^{\mu}\partial_{\mu}-m_{0}c\right]\psi=0,\label{Cdirac}
\end{equation}

where: $\Gamma^{\mu}=\gamma^{\mu}A^{\mu}$ and $A^{\mu}$ is the electromagnetic field of the particle (see Nyambuya $2007, 2008a,b$). So the question is: ``Given that $m_{0}c=\epsilon Q$ and that $A^{\mu}$ is the electromagnetic field of the particle, and that the Dirac Equation is a special case of (\ref{Cdirac}) where $|A^{\mu}|=1$;  what then, is the complete package of transformation for the Dirac Equation that comes along with the reversal of the particle's electromagnetic properties?'' A reversal of the particles electromagnetic properties means we have to reverse the field $A^{\mu}$ and the particles electronic charge $Q$, i.e. $A^{\mu}\longmapsto -A^{\mu}$ and the particles electronic charge $Q\longmapsto -Q$. The transformation: $A^{\mu}\longmapsto -A^{\mu}\Longrightarrow \Gamma^{\mu}\longmapsto -\Gamma^{\mu}$. In flat spacetime as in the case of the Dirac Equation, $|A^{\mu}|=1$ $\Longrightarrow$ $\Gamma^{\mu}=\gamma^{\mu}$. What all this means for the Dirac Equation is that, in the event that we reverse the electromagnetic properties of the particle, the $\gamma$-matrices will transform: $\gamma^{\mu}\longmapsto -\gamma^{\mu}$, that is:

\begin{equation}
Q\longmapsto-Q\Longrightarrow \gamma^{\mu}\longmapsto -\gamma^{\mu},\label{ctrans}
\end{equation}

is the complete package of transformation for the Dirac Equation that comes along with the reversal of the particle's electromagnetic properties.

Now, we proceed to derive the sort for equation in which the Dirac Equation is endowed with a cosmological field in the time dimension. We shall consider the two avenues which to arrive at equation (\ref{Emc4}) separately and in the end, put forward a reason for rejecting one over the other.

\subsection{\subsectionfont Case I}

Given the Dirac Equation, the transformation:

\begin{equation}
\frac{\partial}{\partial t}\longrightarrow \frac{\partial}{\partial t} \pm i\Lambda_{0} c,
\end{equation}

leads us to the modified Dirac Equation, namely:

\begin{equation}
i\hbar\gamma^{\mu}\partial_{\mu}\psi\pm\Lambda_{0} \hbar \gamma^{0}\psi=\epsilon Q\psi.\label{cdirac}
\end{equation}

Let us call this the Partial Cosmological Dirac Equation and the reason for calling it the ``Partial Cosmological'' Dirac Equation is because this equation, unlike the original Dirac Equation -- contains a term ($\Lambda_{0}$) of cosmological significance and this is partial because the cosmological field is confined to the time dimension ($x^{0}$) and not the space dimensions ($x^{1}, x^{2}, x^{3}$) -- for this reason, it is only partial in that it does not cover all the $4$-dimensions of spacetime ($x^{0}, x^{1}, x^{2}, x^{3}$). 

Unlike the original Dirac Equation, this equation possesses no perfect symmetry but:

\begin{enumerate}
\item Violates $\rm{C}$-\textit{symmetry}.

\item Obeys $\rm{T}$-\textit{symmetry}.

\item Obeys $\rm{P}$-\textit{symmetry}.

\item Violates $\rm{CT}$-\textit{symmetry}.

\item Violates $\rm{CP}$-\textit{symmetry}.

\item Obeys $\rm{PT}$-\textit{symmetry}.

\item Violates $\rm{CPT}$-\textit{symmetry}.

\item Obeys Lorentz invariance.
\end{enumerate}

We shall demonstrate these symmetries for equation (\ref{cdirac}) and, as a word of caution, we hope the reader does not get de-focused from the main theme of this reading, namely that we would like to show that the inclusion of a $4$-vector cosmological field does in principle, explain the existing asymmetry between matter and antimatter. Actually, this $4$-vector cosmological field explains more than the this as will be seen. In this exercise, we shall consider the case $+\Lambda_{0}$ and this clearly proves the case for $-\Lambda_{0}$.

\subsubsection{\subsubsectionfont C-Symmetry\label{c}} 

To show invariance  under charge conjugation (or lack thereof), we proceed as usual -- by bringing the particle under the influence of an external electromagnetic magnetic field $A_{\mu}^{ex}$ (which is a real function and this is the usual four vector electromagnetic potential) which leads to the transformation: $\partial_{\mu} \longrightarrow \textrm{D}_{\mu}=\partial_{\mu}+iA_{\mu}^{ex}$, hence equation (\ref{cdirac}) will now be given by: 

\begin{equation}
i\hbar\gamma^{\mu}\rm{D}_{\mu}\psi+\Lambda_{0} \hbar \gamma^{0}\psi=\epsilon Q\psi.\label{cinv1}
\end{equation}

Now, if the equation is invariant under charge conjugation, then, equation (\ref{cinv1}) must under a reversal of the external electromagnetic field ($A_{\mu}^{ex}\longmapsto-A_{\mu}^{ex}\Longrightarrow\rm{D}_{\mu}\longmapsto\rm{D}^{*}_{\mu}$ where the asterisk represents, as usual, the  complex conjugate) and that of the particle ($Q\longmapsto-Q$ \& $\gamma^{\mu}\longmapsto-\gamma^{\mu}$ -- remember \ref{ctrans}),  revert back to the original equation (\ref{cdirac}) and this after a set of transformations of the spinor field $\psi$. Thus reversal of the ambient electromagnetic field and that of the particle leads equation (\ref{cinv1})  to be given by: $-i\hbar\gamma^{\mu}\rm{D}_{\mu}^{*}\psi-\Lambda_{0} \hbar \gamma^{0}\psi=-\epsilon Q\psi$.  Now, to revert back to the original equation, we begin by taking the complex conjugate on both sides of this equation,  that is: $+i\hbar\gamma^{\mu*}\rm{D}_{\mu}\psi^{*}-\Lambda_{0} \hbar \gamma^{0*}\psi^{*}=-\epsilon Q\psi^{*}$. Further, we multiply both sides of this equation by $\gamma^{2}$, and we are lead to: $+i\hbar\gamma^{2}\gamma^{\mu*}\rm{D}_{\mu}\psi^{*}-\Lambda_{0} \hbar \gamma^{2}\gamma^{0*}\psi^{*}=-\epsilon Q\gamma^{2}\psi^{*}$, and using the fact that: $\gamma^{2}\gamma^{\mu}=-\gamma^{\mu}\gamma^{2}$ for $\,\mu\neq2$, and that: $\gamma^{1*}=\gamma^{1}$, $\gamma^{2*}=-\gamma^{2}$ and: $\gamma^{3*}=\gamma^{3}$, we are lead to: $-i\hbar\gamma^{\mu}\rm{D}_{\mu}\psi_{c}+\Lambda_{0} \hbar\gamma^{0}\psi_{c}=-\epsilon Q\psi_{c}$ and multiplying this throughout  by $-1$, we will have:

\begin{equation}
i\hbar\gamma^{\mu}\rm{D}_{\mu}\psi_{c}-\Lambda_{0} \hbar\gamma^{0}\psi_{c}=\epsilon Q\psi_{c},\label{cinv2}
\end{equation}

where: $\psi_{c}=\gamma^{2}\psi^{*}$.  To  revert back to the original equation [which is equation (\ref{cdirac})], the sign in the second term on the right-handside must be positive and for this to be so, one would need a $4\times4$ matrix $M$ such that: $M\gamma^{0}=-\gamma^{0}M$ and: $M\gamma^{\mu}=\gamma^{\mu}M$, and the only matrix  satisfying these conditions is the null matrix. Clearly, multiplying by a null matrix is meaningless hence thus this equation \textbf{\underline{violates $\rm{C}$-\textit{symmetry}}}. 

\subsubsection{\subsubsectionfont T-Symmetry\label{t}} 

To show invariance (or lack thereof) under time reversal, we proceed as usual -- by making the transformation: $t\longmapsto-t$ ($\Longrightarrow \partial_{0}\longmapsto-\partial_{0}$) into (\ref{cdirac}) resulting in this equation reducing to: $-i\hbar\gamma^{0}\partial_{0}\psi+i\hbar\gamma^{k}\partial_{k}\psi+\Lambda_{0} \hbar \gamma^{0}\psi=\epsilon Q\psi$ (NB, $k=1,2,3$). Taking the complex conjugate and then multiplying this equation by $\gamma^{5}\gamma^{2}$ and using the fact that: $\gamma^{2}\gamma^{\mu}=-\gamma^{\mu}\gamma^{2}$ for $\,\mu\neq2$, $\gamma^{5}\gamma^{\mu}=-\gamma^{\mu}\gamma^{5}$, $\gamma^{1*}=\gamma^{1}$, $\gamma^{2*}=-\gamma^{2}$ and $\gamma^{3*}=\gamma^{3}$, one is lead to the original equation (\ref{cdirac}): $i\hbar\gamma^{\mu}\partial_{\mu}\psi_{c}+\Lambda_{0} \hbar \gamma^{0}\psi_{c}=\epsilon Q\psi_{c}$ where: $\psi_{c}=\gamma^{5}\gamma^{2}\psi^{*}$, hence thus equation (\ref{cdirac})  \textbf{\underline{is $\rm{T}$\textit{-symmetric}}}.

\subsubsection{\subsubsectionfont P-Symmetry\label{p}} 

To show invariance under space reversal (or lack thereof) , we proceed as usual -- by making the transformation: $x^{k}\longmapsto-x^{k}$ ($\Longrightarrow \partial_{k}\longmapsto-\partial_{k}$) into (\ref{cdirac}) resulting in this equation reducing to: $i\hbar\gamma^{0}\partial_{0}\psi-i\hbar\gamma^{k}\partial_{k}\psi+\Lambda_{0} \hbar \gamma^{0}\psi=\epsilon Q\psi$. Now to revert back to the original equation,  we simple multiplying both sides of this equation by $\gamma^{2}$  and then using the fact that: $\gamma^{2}\gamma^{k}=-\gamma^{k}\gamma^{2}$, one is lead to: $i\hbar\gamma^{0}\partial_{0}\psi_{c}+i\hbar\gamma^{k}\partial_{k}\psi_{c}+\Lambda_{0} \hbar \gamma^{0}\psi_{c}=\epsilon Q\psi_{c}$ where: $\psi_{c}=\gamma^{2}\psi$. This equation is the same as (\ref{cdirac}) hence equation (\ref{cdirac}) \textbf{\underline{is $\rm{P}$\textit{-symmetric}}}.

\subsubsection{\subsubsectionfont CT-Symmetry\label{ct}} 

To show invariance  under charge conjugation and time reversal (or lack thereof), we proceed by making the transformation:  $\partial_{\mu} \longrightarrow \textrm{D}_{\mu}$ (for the introduction of the external electromagnetic field), $Q\longmapsto-Q$ \& $\gamma^{\mu}\longmapsto-\gamma^{\mu}$ (for the reversal of the particle's electromagnetic properties -- remember \ref{ctrans}), and $t\longmapsto-t$ ($\Longrightarrow \partial_{0}\longmapsto-\partial_{0}$) (for the reversal of time) into (\ref{cdirac}) and this results in this equation reducing to: $-i\hbar\gamma^{0}\partial_{0}\psi-\hbar\gamma^{0}A_{0}^{ex}\psi+i\hbar\gamma^{k}\rm{D}_{k}\psi+\Lambda_{0} \hbar \gamma^{0}\psi=-\epsilon Q\psi$. Now, reversing the external electromagnetic field and taking the complex conjugate both sides, we are lead to: $i\hbar\gamma^{0*}\partial_{0}\psi^{*}+\hbar\gamma^{0*}A_{0}^{ex}\psi-i\hbar\gamma^{k*}\rm{D}_{k}\psi^{*}+\Lambda_{0} \hbar \gamma^{0*}\psi^{*}=-\epsilon Q\psi^{*}$. Multiplying both sides of this equation by $\gamma^{2}$,  and then using the fact that: $\gamma^{2}\gamma^{\mu}=-\gamma^{\mu}\gamma^{2}$ for $\,\mu\neq2$ and that: $\gamma^{1*}=\gamma^{1}$, $\gamma^{2*}=-\gamma^{2}$ and $\gamma^{3*}=\gamma^{3}$, we obtain: $-i\hbar\gamma^{0}\partial_{0}\psi_{c}-\hbar\gamma^{0}A_{0}^{ex}\psi_{c}+i\hbar\gamma^{k}\rm{D}_{k}\psi_{c}-\Lambda_{0} \hbar \gamma^{0}\psi_{c}=-\epsilon Q\psi_{c}$, and now multiplying this by $\gamma^{0}$,  and using the fact that: $\gamma^{0}\gamma^{k}=-\gamma^{k}\gamma^{0}$, we will have: $-i\hbar\gamma^{0}\partial_{0}\psi_{c}-\hbar\gamma^{0}A_{0}^{ex}\psi_{c}-i\hbar\gamma^{k}\rm{D}_{k}\psi_{c}-\Lambda_{0} \hbar \gamma^{0}\psi_{c}=-\epsilon Q\psi_{c}$, where: $\psi=\gamma^{0}\gamma^{2}\psi^{*}$. Multiplying this throughout by $-1$, one is lead to: $i\hbar\gamma^{0}\partial_{0}\psi_{c}+\hbar\gamma^{0}A_{0}^{ex}\psi_{c}+i\hbar\gamma^{k}\rm{D}_{k}\psi_{c}+\Lambda_{0} \hbar \gamma^{0}\psi_{c}=\epsilon Q\psi_{c}$. Now, for this equation to be the same as equation (\ref{cdirac}), we would need the second term on the left handside to change its sign while all other terms retain their signs -- there is no operation in existence that can do this, hence thus equation (\ref{cdirac}) \textbf{\underline{violates $\rm{CT}$-\textit{symmetry}}}.

\subsubsection{\subsubsectionfont CP-Symmetry\label{cp}} 

To show invariance  under charge conjugation and space reversal (or lack thereof), we proceed  by making the transformation:  $\partial_{\mu} \longrightarrow \textrm{D}_{\mu}$ (for the introduction of the electromagnetic field), $Q\longmapsto-Q$ \& $\gamma^{\mu}\longmapsto-\gamma^{\mu}$ (for the reversal of the electromagnetic properties of the particle -- remember \ref{ctrans}) and  $x^{k}\longmapsto-x^{k}$  ($\Longrightarrow\partial_{k}\longmapsto-\partial_{k}$) (for the reversal of the space coordinates) into (\ref{cdirac}), one is lead to: $-i\hbar\gamma^{0}\partial_{0}\psi+\hbar\gamma^{0}A_{0}\psi-i\hbar\gamma^{k}\rm{D}_{k}^{*}\psi-\Lambda_{0} \hbar \gamma^{0}\psi=-\epsilon Q\psi$. Now, reversing the external electromagnetic field and taking the complex conjugate both sides, we are lead to: $+i\hbar\gamma^{0}\partial_{0}\psi-\hbar\gamma^{0}A_{0}\psi+i\hbar\gamma^{k*}\rm{D}_{k}\psi^{*}-\Lambda_{0} \hbar \gamma^{0*}\psi^{*}=-\epsilon Q\psi^{*}$. Multiplying both sides of this equation by $\gamma^{2}$,  and then using the fact that: $\gamma^{2}\gamma^{\mu}=-\gamma^{\mu}\gamma^{2}$ for $\,\mu\neq2$ and that: $\gamma^{1*}=\gamma^{1}$, $\gamma^{2*}=-\gamma^{2}$ and $\gamma^{3*}=\gamma^{3}$, we obtain: $\gamma^{0}\gamma^{k}=-\gamma^{k}\gamma^{0}$, we obtain: $+i\hbar\gamma^{0}\partial_{0}\psi_{c}-\hbar\gamma^{0}A_{0}\psi_{c}+i\hbar\gamma^{k}\rm{D}_{k}\psi_{c}-\Lambda_{0} \hbar \gamma^{0}\psi_{c}=\epsilon Q\psi_{c}$ where $\psi_{c}=-\gamma^{2}\psi^{*}$. Now, for this equation to be the same as equation (\ref{cdirac}), we would need the last term on the left handside to change its sign while all other terms retain their signs -- there is no operation in existence that can do this, hence thus equation (\ref{cdirac}) \textbf{\underline{violates $\rm{CP}$-\textit{symmetry}}}.

\subsubsection{\subsubsectionfont PT-Symmetry\label{pt}}

To show invariance  under a combined space and time reversal (or lack thereof), we proceed  by making the transformation: $x^{\mu}\longmapsto-x^{\mu}$ ($\Longrightarrow\partial_{\mu}\longmapsto-\partial_{\mu}$ and this is for the reversal of the spacetime coordinates) into (\ref{cdirac}), we are lead to:  $-i\hbar\gamma^{\mu}\partial_{\mu}\psi+\Lambda_{0} \hbar \gamma^{0}\psi=\epsilon Q\psi$. To  revert back to the original equation (\ref{cdirac}), we take the complex conjugate both-sides and then multiply throughout by $\gamma^{0}\gamma^{2}\gamma^{5}$, and we are lead to: $i\hbar\gamma^{\mu}\partial_{\mu}\psi_{c}+\Lambda_{0} \hbar \gamma^{0}\psi_{c}=\epsilon Q\psi_{c}$, where $\psi_{c}=\gamma^{0}\gamma^{2}\gamma^{5}\psi^{*}$, hence thus equation (\ref{cdirac})  \textbf{\underline{is $\rm{PT}$-\textit{symmetry}}}. 

\subsubsection{\subsubsectionfont CPT-Symmetry\label{cpt}} 

To show invariance  under a combined charge, space and time reversal (or lack thereof), we proceed  by making the transformation:  $\partial_{\mu} \longrightarrow \textrm{D}_{\mu}$ (for the introduction of the external electromagnetic field), $Q\longmapsto-Q$ \& $\gamma^{\mu}\longmapsto-\gamma^{\mu}$ (for the reversal of the electromagnetic properties of the particle -- remember \ref{ctrans}) and $x^{\mu}\longmapsto-x^{\mu}$ ($\Longrightarrow\partial_{\mu}~\longrightarrow~-\partial_{\mu}$ and this for the reversal of the space and time coordinates) into (\ref{cdirac}) and this leads us to: $+i\hbar\gamma^{\mu}\partial_{\mu}\psi-i\hbar\gamma^{\mu}A_{\mu}^{ex}\psi-\Lambda_{0} \hbar \gamma^{0}\psi=-\epsilon Q\psi$. Now, reversing the external electromagnetic field, we are lead to: $i\hbar\gamma^{\mu}\rm{D}_{\mu}\psi-\Lambda_{0} \hbar\gamma^{0}\psi=-\epsilon Q\psi$. Just as in the case of the calculation of $\rm{C}$-\textit{symmetry} in \$ (\ref{c}), to  revert back to the original equation (\ref{cdirac}), the sign in the first term on the right-handside must be negative so that the new equation would read: $i\hbar\gamma^{\mu}\rm{D}_{\mu}\psi_{c}+\Lambda_{0} \hbar\gamma^{0}\psi_{c}=\epsilon Q\psi_{c}$ where $\psi_{c}=-\psi$, and for this to be so; one would need a $4\times4$ matrix $M$ such that: $M\gamma^{0}=\gamma^{0}M$ \& $M\gamma^{\mu}=-\gamma^{\mu}M$, and the only matrix  satisfying these conditions is the null matrix  hence thus this equation \textbf{\underline{violates $\rm{CPT}$-\textit{symmetry}}}. 

\subsubsection{\subsubsectionfont Lorentz Invariance\label{l}} 

To prove Lorentz invariance for (\ref{cdirac}) (which we shall write as: $[i\hbar\gamma^{\mu}\partial_{\mu}-\Lambda_{0} \hbar \gamma^{0}-\epsilon Q]\psi=0$), two conditions must be satisfied:

1. Given any two inertial observers $\textrm{O}$ and $\textrm{O}^\prime$ anywhere in spacetime, if in the frame $\textrm{O}$ we have $[i\hbar\gamma^{\mu}\partial_{\mu}-\Lambda_{0} \hbar \gamma^{0}-\epsilon Q]\psi(x)=0$, then: $[i\hbar\gamma^{\mu\prime}\partial_{\mu}^{\prime}-\Lambda_{0}^{\prime} \hbar \gamma^{0\prime}-\epsilon Q]\psi^\prime(x^\prime)=0$, is the equation describing the same state but in the frame $\textrm{O}^\prime$. 

2. Given that $\psi(x)$ is the wavefunction as measured by observer $\textrm{O}$, there must be a prescription for observer $\textrm{O}^\prime$ to compute $\psi^\prime(x^\prime)$ from $\psi(x)$ and this describes to $\textrm{O}^\prime$ the same physical state as that measured by $\textrm{O}$.

Now, since the Lorentz transformation are linear, it is to be required or expected of the transformations between $\psi(x)$ and $\psi^\prime(x^\prime)$  to be linear too, that is:

\begin{equation} 
\psi^\prime(x^\prime) = \psi^\prime(\Gamma x) = S(\Gamma) \psi(x) = S(\Gamma) \psi(\Gamma^{-1}x^\prime)\label{inverse1}
\end{equation} 
\\
where $S(\Gamma)$ is a $4\times 4$ matrix which depends only on the relative velocities of $\textrm{O}$ and $\textrm{O}^\prime$ and $\Gamma$ is the Lorentz transformation matrix. $S(\Gamma)$ has an inverse if $\textrm{O}\rightarrow \textrm{O}^\prime$ and also $\textrm{O}^\prime\rightarrow \textrm{O}$. The inverse is:

\begin{equation} 
\psi(x) = S^{-1}(\Gamma)\psi^\prime(x^\prime) = S^{-1}(\Gamma)\psi^\prime(\Gamma x) \label{inverse2}
\end{equation} 	

or we could write:

\begin{equation}
\psi(x)=S(\Gamma^{-1})\psi^\prime(\Gamma x)\Longrightarrow S(\Gamma^{-1}) = S^{-1}(\Gamma)
\end{equation}

We can now write: $[i\hbar\gamma^{\mu}\partial_{\mu}-\Lambda_{0} \hbar \gamma^{0}-\epsilon Q]\psi(x)=0$, 
as:  $[i\hbar\gamma^{\mu}\partial_{\mu}-\Lambda_{0} \hbar \gamma^{0}-\epsilon Q]S^{-1}(\Gamma)\psi^\prime(x^\prime)=0$, and multiplying this from the left by $S(\Gamma)$, we have: $S(\Gamma)[i\hbar\gamma^{\mu}\partial_{\mu}-\Lambda_{0} \hbar \gamma^{0}-\epsilon Q] S^{-1}(\Gamma)\psi^\prime(x^{\prime})=0$,
and hence:

\begin{equation}
\left[i\hbar S(\Gamma)\gamma^{\mu} S^{-1}(\Gamma)\partial_\mu -\Lambda_{0} \hbar S(\Gamma)\gamma^{0} S^{-1}(\Gamma)-\epsilon Q\right]\psi^\prime(x^\prime)=0.\label{lorenz}
\end{equation}

Now, the reader must take note of the fact that the cosmological  field is a vector quantity and transforms as:

\begin{equation}
\Lambda_{0}=\left(\frac{\partial x^{0\prime}}{\partial x^{0}}\right)\Lambda_{0\prime}. \label{l-trans}
\end{equation}

The reason for this is, that, it is a property of time and thus it will have to transforms in the same manner as time does. For better clarity: $(\partial_{0}+\Lambda_{0})=(\partial x^{0\prime}/\partial x^{0})(\partial_{0}+\Lambda_{0})^{\prime}$, and from this -- equation (\ref{l-trans}) smoothly flows. Given this, and also that: $\partial_{\mu}=(\partial x^{\mu\prime}/\partial x^{\mu})\partial_{\mu\prime}$, and putting all this into (\ref{lorenz}), we are lead to:

\begin{widetext}
\begin{equation}
\left[i\hbar \left(\frac{\partial x^{\mu\prime}}{\partial x^{\mu}}\right) S(\Gamma)\gamma^{\mu} S^{-1}(\Gamma)\partial_\mu^\prime -\Lambda_{0}^\prime\hbar\left(\frac{\partial x^{0\prime}}{\partial x^{0}}\right) S(\Gamma)\gamma^{0} S^{-1}(\Gamma)-\epsilon Q\right]\psi^\prime(x^\prime)=0.
\label{lorenz2}
\end{equation}
\end{widetext}

Now making the setting:

\begin{equation} 
\gamma^{\mu\prime}=\left(\frac{\partial x^{\mu\prime}}{\partial x^{\mu}}\right)S(\Gamma)\gamma^{\mu} S^{-1}(\Gamma), 
\end{equation}

equation (\ref{lorenz2}) can now be written as: 

\begin{equation}
\left[i\hbar \gamma^{\mu\prime}\partial_\mu^\prime -\Lambda_{0}^\prime\hbar\gamma^{0\prime}-\epsilon Q\right]\psi^\prime(x^\prime)=0,
\end{equation}

hence thus equation (\ref{cdirac}) \textbf{\underline{is Lorentz invariant}} thus satisfying one of the necessary requirements for this equation to be physically meaningful. Now -- the question is, does this equation have all it takes to have correspondence with physical reality? To answer this question, we shall inspect this equation's Hamiltonian.

\subsubsection{\subsubsectionfont Hamiltonian}

Now that we have investigated the symmetries of equation (\ref{cdirac}), the question is \textit{``Does this equation qualify -- in principle, to describe physical phenomena?''} The answer is yes and first and forest; despite it's violation of $\rm{CPT}$-\textit{symmetry}, it is Lorentz invariant and second; because its Hamiltonian, namely:

\begin{equation}
\mathcal{H}=-iI\hbar\frac{\partial }{c\partial t}=i\hbar\gamma^{0}\gamma^{k}\partial_{k}\pm \Lambda_{0}\hbar I - \epsilon Q\gamma^{0},
\end{equation}

(where $I$ is and hereafter the $4\times4$ identity matrix) is hermitian (one can easily verify this for themself, hermiticity means: $\mathcal{H}^{\dagger}=\mathcal{H}$) means its the energy eigenvalues are real. One could argue that this equation be rejected on the grounds that it violates a cornerstone theorem of Quantum Field Theory (QFT), namely Schwinger-L\"uder-Pauli theorem also known as the $\rm{CPT}$ theorem (see e.g. Greaves $2007$ for a good exposition). The L\"uder-Pauli theorem states that any Lorentz invariant local QFT with a Hermitian Hamiltonian must possess or obey $\rm{CPT}$-\textit{symmetry}. This theorem is derived for a symmetric spacetime and not a none-symmetric spacetime as the present hence thus it does not apply here. We shall address this issue fully in \S (\ref{cptv_s}).

\subsection{\subsectionfont Case II}

Proceeding to the second case, we make the following transformation:

\begin{equation}
\frac{\partial}{\partial t}\longrightarrow \frac{\partial}{\partial t} \pm\Lambda_{0} c,
\end{equation}

and this leads us to:

\begin{equation}
i\hbar\gamma^{\mu}\partial_{\mu}\psi\pm i\Lambda_{0}\hbar \gamma^{0}\psi=\epsilon Q\psi.\label{cdirac2}
\end{equation}

As one can verify for themself, this equation possesses the following symmetries:

\begin{enumerate}
\item Violates $\rm{C}$-\textit{symmetry}.

\item Violates $\rm{T}$-\textit{symmetry}.

\item Obeys P-\textit{symmetry}.

\item Violates $\rm{CT}$-\textit{symmetry}.

\item Violates $\rm{CP}$-\textit{symmetry}.

\item Violates $\rm{PT}$-\textit{symmetry}.

\item Violates $\rm{CPT}$-\textit{symmetry}.

\item Obeys Lorentz invariance.
\end{enumerate}

Using the same methods as shown in \S (\ref{c}) to \S (\ref{l}), one can demonstrate and verify for themselves that indeed, (\ref{cdirac2}) exhibits the above said symmetries. Now the question is, what is the relationship of this equation with physical reality? To answer this question, we shall inspect this equation's Hamiltonian.

\subsubsection{\subsubsectionfont Hamiltonian}

Clearly -- at the very least, equation (\ref{cdirac2}) is in contempt of physical reality since its Hamiltonian, namely:

\begin{equation}
\mathcal{H}=i\hbar\gamma^{0}\gamma^{k}\partial_{k}-i\Lambda_{0} \hbar I-\epsilon Q\gamma^{0},
\end{equation}

is (as one can verify for themselves) not hermitian thus leads to complex energy eigenvalues. Complex energy eigenvalues are physically meaningless hence thus this equation ought to be rejected outright with the simple remark that ``it has no bearing with physical reality as we know it.''  

\section{\sectionfont The Arrow of Time\label{at}}

Now having gone through the symmetries of equation (\ref{cdirac}) and having chosen this equation and rejected equation (\ref{cdirac2}), we come to the question of the arrow of time and the results obtained here extend to the equation that we will derive in the next section. We note here -- that, we have two Universes, one described by $-\Lambda_{0}$ and the other by $+\Lambda_{0}$. The question is, what is the arrow of time in these two Universes? Is it directed in the forward or in the backward direction? Within the provinces of the present theory, this question can be answered if we combined the present with one of the basic principles of QM -- namely that the wavefunction ought to be normalizable. To reach this end, we shall solve for the free particle solution for equation (\ref{cdirac}) and inspect these and accept only those solutions that ``behave''. Let -- as is usual:  $\psi=u_{p}e^{ip_{\mu}x^{\mu}/\hbar}$, where:

\begin{equation}
u_{p} =
\left(\begin{array}{c}
\Phi\\
\\
\chi
\end{array}
\right),
\,\,
\rm{and \, where:} 
\,\, 
\begin{array}{c c}
\Phi = \left(\begin{array}{c}
\Phi_{1}\\
\\
\Phi_{2}
\end{array}
\right) & \,\,\rm{and}\,\,\chi = \left(\begin{array}{c}
\chi_{1}\\
\\
\chi_{2}
\end{array}
\right)
\end{array}.
\end{equation}

At this moment, a very important point to remember is that $p_{0}$ now contains the cosmological field, the meaning of which is that we must write: $p_{0}\longmapsto p_{0}\pm i\Lambda_{0}\hbar c$, thus: $\psi=u_{p}e^{\pm\Lambda_{0}\hbar ct}e^{ip_{\mu}x^{\mu}/\hbar}$. Now, if the probability density function: $\rho(t)=\psi^{\dagger}\psi=u^{\dagger}_{p}u_{p}e^{\pm\Lambda ct}$, is to be finite  as: $t\longmapsto+\infty$ or $t\longmapsto-\infty$ (according to the basic principles of QM, this is prerequisite for any wavefunction), it is evidently clear that for the case  $-\Lambda_{0}$, this will only be so if: $t>0$ [$\rho(t~\longmapsto~\infty)\longmapsto0$] and for the case $+\Lambda_{0}$, we must have: $t<0$ [$\rho(t~\longmapsto~-~\infty)~\longmapsto0$]. Hence thus, the arrow of time in the two Universes is different and moves in opposite directions. 

From the above simple calculation, we conclude that the Universe for which: $\Lambda_{0}<0$, time moves in the forward direction and likewise, the Universe for which: $\Lambda_{0}>0$, the arrow of time moves in the backward direction, hence thus we have to different Universes. Let these two Universes be $\mathcal{U}^{+}$ and   $\mathcal{U}^{-}$ where $\mathcal{U}^{+}$ is the Universe in which time moves forward ($\Lambda_{0}<0$) and likewise, $\mathcal{U}^{-}$  is the Universe in which time moves backwards ($\Lambda_{0}>0$). This result is independent of the the fact that we have here considered a free particle as it will hold true for all conditions of experience.

\section{\sectionfont Full  Cosmological Dirac Equation}

Given the unquenchable thirst to generalise Laws \textit{of} Nature, it is most natural to wonder:  ``Why should the cosmological  field be confined just to the time dimension alone?'' Why not the space dimension as-well? To introduce a cosmological  field into a particular dimension, one simply needs to add this to the partial derivative of that dimension, thus to have this in all the four dimensions we have to perform the transformation:

\begin{equation}
\partial_{\mu}\longmapsto\partial_{\mu}+\Lambda_{\mu}, 
\end{equation}

where:

\begin{equation}
\Lambda_{\mu}\equiv[\omega_{0}i\Lambda_{0}, \omega_{1}\Lambda_{1}, \omega_{2}\Lambda_{2}, \omega_{3}\Lambda_{3}],\label{cosm_const}
\end{equation}

and: $\omega_{\mu}=\pm1$ ($\omega_{\mu}$ is not a four vector but a simple number). In this way, i.e. -- the addition of the cosmological field in the space dimensions, we have endowed the vacuum of space with some all-pervading and permeating momentum. This modification (equation \ref{cosm_const}) automatically  leads to the  $4$-momentum transforming:  $p^{k}\longrightarrow \mathcal{P}_{k}= p_{k}+\omega_{k}\Lambda_{k}\hbar$, and plucking this into (\ref{Emc3}), one is lead to:

\begin{equation}
\mathcal{E}=\omega_{0}\Lambda_{0} \hbar c\pm\sqrt{\mathcal{P}^{k}\mathcal{P}_{k}c^{2}+\epsilon ^{2}Q^{2}c^{2}}.\label{cosm_eqn}
\end{equation}

Considering the case $+\Lambda_{0}$, this new energy equation (\ref{cosm_eqn}) interestingly has $\textbf{16}+1=17$ energy solutions! Let these energy solutions be: $\mathcal{E}_{j}$ where $j=1,2,3\, ...\, 16,17$. Explicitly these solutions are given:

\begin{widetext}
{
\begin{equation}
\begin{array}{c c l l l}
\mathcal{E}_{1}& = &+\Lambda_{0} \hbar c+\sqrt{[( p_{1}+\Lambda_{1}\hbar)\cdot(p^{1}+\Lambda^{1}\hbar)+( p_{2}+\Lambda_{2}\hbar)\cdot(p^{2}+\Lambda^{2}\hbar)+( p_{3}+\Lambda_{3}\hbar)\cdot(p^{3}+\Lambda^{3}\hbar)]c^{2}+\epsilon ^{2}Q^{2}c^{2}}& ... & (\textbf{a})\\
\\
\mathcal{E}_{2}& = &+\Lambda_{0} \hbar c+\sqrt{[( p_{1}-\Lambda_{1}\hbar)\cdot(p^{1}-\Lambda^{1}\hbar)+( p_{2}+\Lambda_{2}\hbar)\cdot(p^{2}+\Lambda^{2}\hbar)+( p_{3}+\Lambda_{3}\hbar)\cdot(p^{3}+\Lambda^{3}\hbar)]c^{2}+\epsilon ^{2}Q^{2}c^{2}}& ... & (\textbf{b})\\
\\
\mathcal{E}_{3}& = &+\Lambda_{0} \hbar c+\sqrt{[( p_{1}+\Lambda_{1}\hbar)\cdot(p^{1}+\Lambda^{1}\hbar)+( p_{2}-\Lambda_{2}\hbar)\cdot(p^{2}-\Lambda^{2}\hbar)+( p_{3}+\Lambda_{3}\hbar)\cdot(p^{3}+\Lambda^{3}\hbar)]c^{2}+\epsilon ^{2}Q^{2}c^{2}}& ... & (\textbf{d})\\
\\
\mathcal{E}_{4}& = &+\Lambda_{0} \hbar c+\sqrt{[( p_{1}+\Lambda_{1}\hbar)\cdot(p^{1}+\Lambda^{1}\hbar)+( p_{2}+\Lambda_{2}\hbar)\cdot(p^{2}+\Lambda^{2}\hbar)+( p_{3}-\Lambda_{3}\hbar)\cdot(p^{3}-\Lambda^{3}\hbar)]c^{2}+\epsilon ^{2}Q^{2}c^{2}}& ... & (\textbf{e})\\
\\
\mathcal{E}_{5}& = &+\Lambda_{0} \hbar c+\sqrt{[( p_{1}+\Lambda_{1}\hbar)\cdot(p^{1}+\Lambda^{1}\hbar)+( p_{2}-\Lambda_{2}\hbar)\cdot(p^{2}-\Lambda^{2}\hbar)+( p_{3}-\Lambda_{3}\hbar)\cdot(p^{3}-\Lambda^{3}\hbar)]c^{2}+\epsilon ^{2}Q^{2}c^{2}}& ... & (\textbf{g})\\
\\
\mathcal{E}_{6}& = &+\Lambda_{0} \hbar c+\sqrt{[( p_{1}-\Lambda_{1}\hbar)\cdot(p^{1}-\Lambda^{1}\hbar)+( p_{2}+\Lambda_{2}\hbar)\cdot(p^{2}+\Lambda^{2}\hbar)+( p_{3}-\Lambda_{3}\hbar)\cdot(p^{3}-\Lambda^{3}\hbar)]c^{2}+\epsilon ^{2}Q^{2}c^{2}}& ... & (\textbf{h})\\
\\
\mathcal{E}_{7}& = &+\Lambda_{0} \hbar c+\sqrt{[( p_{1}-\Lambda_{1}\hbar)\cdot(p^{1}-\Lambda^{1}\hbar)+( p_{2}-\Lambda_{2}\hbar)\cdot(p^{2}-\Lambda^{2}\hbar)+( p_{3}+\Lambda_{3}\hbar)\cdot(p^{3}+\Lambda^{3}\hbar)]c^{2}+\epsilon ^{2}Q^{2}c^{2}}& ... & (\textbf{i})\\
\\
\mathcal{E}_{8}& = &+\Lambda_{0} \hbar c+\sqrt{[( p_{1}-\Lambda_{1}\hbar)\cdot(p^{1}-\Lambda^{1}\hbar)+( p_{2}-\Lambda_{2}\hbar)\cdot(p^{2}-\Lambda^{2}\hbar)+( p_{3}-\Lambda_{3}\hbar)\cdot(p^{3}-\Lambda^{3}\hbar)]c^{2}+\epsilon ^{2}Q^{2}c^{2}}& ... & (\textbf{j})\\
\\
\mathcal{E}_{9}& =   &+\Lambda_{0} \hbar c = \mathcal{E}_{vac} & ... & (\textbf{f})\\
\end{array}
\end{equation}
}

and the negative solutions are:
{
\begin{equation}
\begin{array}{c c l l l}
\mathcal{E}_{10}& = &+\Lambda_{0} \hbar c-\sqrt{[( p_{1}-\Lambda_{1}\hbar)\cdot(p^{1}-\Lambda^{1}\hbar)+( p_{2}-\Lambda_{2}\hbar)\cdot(p^{2}-\Lambda^{2}\hbar)+( p_{3}-\Lambda_{3}\hbar)\cdot(p^{3}-\Lambda^{3}\hbar)]c^{2}+\epsilon ^{2}Q^{2}c^{2}}& ... & (\textbf{j})\\
\\
\mathcal{E}_{11}& = &+\Lambda_{0} \hbar c-\sqrt{[( p_{1}-\Lambda_{1}\hbar)\cdot(p^{1}-\Lambda^{1}\hbar)+( p_{2}-\Lambda_{2}\hbar)\cdot(p^{2}-\Lambda^{2}\hbar)+( p_{3}+\Lambda_{3}\hbar)\cdot(p^{3}+\Lambda^{3}\hbar)]c^{2}+\epsilon ^{2}Q^{2}c^{2}}& ... & (\textbf{i})\\
\\
\mathcal{E}_{12}& = &+\Lambda_{0} \hbar c-\sqrt{[( p_{1}-\Lambda_{1}\hbar)\cdot(p^{1}-\Lambda^{1}\hbar)+( p_{2}+\Lambda_{2}\hbar)\cdot(p^{2}+\Lambda^{2}\hbar)+( p_{3}-\Lambda_{3}\hbar)\cdot(p^{3}-\Lambda^{3}\hbar)]c^{2}+\epsilon ^{2}Q^{2}c^{2}}& ... & (\textbf{h})\\
\\
\mathcal{E}_{13}& = &+\Lambda_{0} \hbar c-\sqrt{[( p_{1}+\Lambda_{1}\hbar)\cdot(p^{1}+\Lambda^{1}\hbar)+( p_{2}-\Lambda_{2}\hbar)\cdot(p^{2}-\Lambda^{2}\hbar)+( p_{3}-\Lambda_{3}\hbar)\cdot(p^{3}-\Lambda^{3}\hbar)]c^{2}+\epsilon ^{2}Q^{2}c^{2}}& ... & (\textbf{g})\\
\\
\mathcal{E}_{14}& = &+\Lambda_{0} \hbar c-\sqrt{[( p_{1}+\Lambda_{1}\hbar)\cdot(p^{1}+\Lambda^{1}\hbar)+( p_{2}+\Lambda_{2}\hbar)\cdot(p^{2}+\Lambda^{2}\hbar)+( p_{3}-\Lambda_{3}\hbar)\cdot(p^{3}-\Lambda^{3}\hbar)]c^{2}+\epsilon ^{2}Q^{2}c^{2}}& ... & (\textbf{e})\\
\\
\mathcal{E}_{15}& = &+\Lambda_{0} \hbar c-\sqrt{[( p_{1}+\Lambda_{1}\hbar)\cdot(p^{1}+\Lambda^{1}\hbar)+( p_{2}-\Lambda_{2}\hbar)\cdot(p^{2}-\Lambda^{2}\hbar)+( p_{3}+\Lambda_{3}\hbar)\cdot(p^{3}+\Lambda^{3}\hbar)]c^{2}+\epsilon ^{2}Q^{2}c^{2}}& ... & (\textbf{d})\\
\\
\mathcal{E}_{16}& = &+\Lambda_{0} \hbar c-\sqrt{[( p_{1}-\Lambda_{1}\hbar)\cdot(p^{1}-\Lambda^{1}\hbar)+( p_{2}+\Lambda_{2}\hbar)\cdot(p^{2}+\Lambda^{2}\hbar)+( p_{3}+\Lambda_{3}\hbar)\cdot(p^{3}+\Lambda^{3}\hbar)]c^{2}+\epsilon ^{2}Q^{2}c^{2}}& ... & (\textbf{b})\\
\\
\mathcal{E}_{17}& = &+\Lambda_{0} \hbar c-\sqrt{[( p_{1}+\Lambda_{1}\hbar)\cdot(p^{1}+\Lambda^{1}\hbar)+( p_{2}+\Lambda_{2}\hbar)\cdot(p^{2}+\Lambda^{2}\hbar)+( p_{3}+\Lambda_{3}\hbar)\cdot(p^{3}+\Lambda^{3}\hbar)]c^{2}+\epsilon ^{2}Q^{2}c^{2}}& ... & (\textbf{a})
\end{array}
\end{equation}
}
\textbf{NB}: The negative and positive energy solutions are symmetric about vacuum energy level $\mathcal{E}_{vac}=\Lambda_{0}\hbar c$.
\\
\end{widetext}

These energies are such that: $\mathcal{E}_{1}\geq\mathcal{E}_{2}\geq ... \geq\mathcal{E}_{16}\geq\mathcal{E}_{17}$ and this is for the setting: $p_{1}\leq p_{2}\leq p_{3}$. The order of which $p$ is greater than the other does not really matter because one can always rearrange these $p$'s in an assending order as has been done here and then  proceed to calculate the energies as above. That said, it is not difficult to see that the spinor field equation describing the energy equation (\ref{cosm_eqn}) is:

\begin{equation}
i\hbar\gamma^{\mu}\partial_{\mu}\psi + i\hbar \gamma^{\mu}\Lambda_{\mu}\psi=\epsilon Q\psi.\label{fsol}
\end{equation}

Let us call this equation the  \textsl{\textbf{Full  Cosmological Dirac Equation}}. We say full because all the four dimensions: $x^{0}-axis$, $x^{1}-axis$, $x^{2}-axis$ and $x^{3}-axis$; are endowed with the cosmological vector field. 

This equation has the following properties:

\begin{enumerate}
\item Violates $\rm{C}$-\textit{symmetry}.

\item Obeys $\rm{T}$-\textit{symmetry}.

\item Violates P-\textit{symmetry}.

\item Violates $\rm{CT}$-\textit{symmetry}.

\item Violates $\rm{CP}$-\textit{symmetry}.

\item Violates $\rm{PT}$-\textit{symmetry}.

\item Violates $\rm{CPT}$-\textit{symmetry}.

\item Obeys Lorentz invariance.
\end{enumerate}

With the aid of the presentations made in \S (\ref{c}) down to \S (\ref{l}), one can verify for themself these symmetries, as doing so [i.e showing these symmetries] here would be a nothing but a reproduction of this same exercise albeit with slight modifications which are not difficult at all.

\section{\sectionfont CPT-violation \& a New Spacetime Model\label{cptv_s}}

Now we come to the problem of $\rm{CPT}$\textit{-violation}. The Schwinger-L\"uder-Pauli theorem, which is a corner stone of QFT, states that any local Lorentz invariant field theory \textbf{\textit{must}} obey $\rm{CPT}$-\textit{symmetry}. $\rm{CPT}$-\textit{symmetry} is thus considered a perfect Symmetry \textit{of} Nature, so much such that any theory that violates it, is not thought to be correct. Despite the strong belief in the preservation of $\rm{CPT}$-\textit{symmetry}, some researchers seeking quantum theories of gravity, take $\rm{CPT}$\textit{-violation} as their point of departure (see e.g. Mavromatos). The equations derived herein are clearly in violation of this \textit{``sacrosanct''} symmetry. Does this mean our ideas are fundamentally incorrect? The answer to this question is -- in our opinion, \textbf{no}! and this is because this theorem applies only to a perfectly symmetric spacetime whereas the spacetime on which the present theory is built is not a symmetric spacetime.

To arrive at the Full Cosmological Dirac Equation (\ref{fsol}), what we have done is to add a constant to the $4$-momentum, that is: $p_{\mu}\longmapsto p_{\mu}+\Lambda_{\mu}\hbar=P_{\mu}$. This suggests that the original spacetime continuum has -- inorder to have the Full Cosmological Dirac Equation (\ref{fsol}), been modified clandestinely, i.e: $x_{\mu}\longmapsto X_{\mu}$. For a particle of whose relativistic mass denoted $m$ and four velocity $\dot{x}_{\mu}$, we know that the four momentum: $p_{\mu}=m\dot{x}_{\mu}$ and in the light of the clandestine modification: $x_{\mu}\longmapsto X_{\mu}$, this implies: $P_{\mu}=m\dot{X}_{\mu}=m\dot{x}_{\mu}+\Lambda_{\mu}\hbar$. Further, this implies: $X_{\mu}=x_{\mu}+(\hbar/m)\int (\Lambda_{\mu}) d\tau$, where $\tau$ is the proper time. Taking $\hbar$ as an absolute fundamental constant and $m$  as having no explicit variation with time, we are left with $\Lambda_{\mu}$ as the only function we can assign a time dependence. 

Given: $X_{\mu}=x_{\mu}+(\hbar/m)\int (\Lambda_{\mu}) d\tau$, we propose that the modified spacetime continuum ($X_{\mu}$) be related to the ordinary spacetime continuum [$x_{\mu}$ -- let us call this spacetime the Classical Spacetime (CST)] by the relationship:

\begin{equation}
X_{\mu}(x)=x_{\mu}+\ell_{p}\delta_{\mu}(t),
\end{equation}

where $\ell_{p}$, is a fundamental absolute constant with the dimensions of length and:

\begin{equation}
\left|\delta_{\mu}(t)\right|\geq1,\label{dirich}
\end{equation}

and $\delta_{\mu}(t)$ is defined on the $x^{\mu}$\textit{-axis}. What is this new function $\delta_{\mu}(t)$, and what is its role? and why does it have the limits its has? We shall give the justification for the limits in \S \ref{stl}.

We propose that this function be a $4$-vector differentiable none-smooth dynamic random function that takes any numerical value within the set limits as defined in (\ref{dirich}). By differentiable none-smooth we mean it is not continuous at any-point but is differentiable at every-point. Such type of functions do exist and where first investigated by the German mathematician Johan P. G. L. Dirichlet (1805-1859). We shall here refer to these functions as Dirichlet Type Functions (DTF). Because the present Dirichlet function has to be random and dynamic, let us call it the Dynamic Random Dirichlet Function (DRDF) and let the spacetime on which it is defined be simple called  Quantum Spacetime (QST) because it has the desired features of  QST, namely:

\begin{enumerate}
\item This spacetime possesses randomness -- randomness is an intrinsic feature of QM. Potentially and very much likely, this randomness may explain the bewildering and mysterious randomness we see in QM.

\item This spacetime, is not itself quantized but it literally quantizes the CST  where it is derived. It quantizes the CST  such that any-two points (no matter how close to each other they may be) on the CST when transformed or cast onto the QST, they will never be closer than the separation $\ell_{p}$ on the $space-axis$ and $t_{p}$ on the time $axis$ on the QST.
\end{enumerate}

This function -- the DTF, is what makes the QST to be intrinsically and inherently $\rm{P}$\textit{-asymmetric} and this is because the DTF is itself not spatially symmetric, that is to say: $\delta_{k}(-x_{k};t)\neq\delta_{k}(+x_{k};t)$ nor $\delta_{k}(-x_{k};t)\neq-\delta_{k}(+x_{k};t)$ and this implies: $X_{k}(-x_{k})\neq-X_{k}(x_{k})$. If: $X_{k}(-x_{k})=-X_{k}(x_{k})$, then the QST would result in $\rm{P}$\textit{-symmetric} laws on this spacetime. 

Said in another manner, in the ordinary $3$-space, the mirror image of the point $x_{k}$ is $-x_{k}$, these points, $x_{k}$ and $-x_{k}$ when cast onto the QST do not transform into mirror images of each other on the QST because the DTF $\delta_{k}(x_{k},t)$, is for every moment in time, unique for every-point. If they where mirror images of each other then: $X_{k}(-x_{k})=-X_{k}(x_{k})$ and as already argued, the nature of: $\delta_{k}(x_{k};t)$ makes this to not hold. 

All said and done, the L\"uder-Pauli $\rm{CPT}$ theorem does not hold in the present case as this has been derived on the CST continuum. The coordinate system of the QST is not symmetric as is the coordinate system of CST $x_{\mu}$, hence thus, any Physical Laws dependent on the coordinates of this coordinates system will be intrinsically and inherently $\rm{P}$\textit{-asymmetric}. The fact that Laws \textit{of} Nature are $\rm{T}$-symmetric on the QST implies: $\delta_{0}(x_{k};t)=\delta_{0}(x_{k};-t)$, and this means the past is preserved by $\delta_{0}$ and is not preserved by $\delta_{k}$. In another way, knowing the present value of $\delta_{0}(x_{k};t)$, means not only can one tell is past value but also foretell its future value. This is interesting and suggests a more rigorous study of the QST.

\subsection{\subsectionfont Upper Space and Time Limits\label{stl}}

In this part of the reading, we establish lower space and time limits on spacetime. To achieve this, we use the simple and well accepted Law \textit{of} Nature that the speed of light, $c$, is an upper absolute speed limit for all material bodies and energy in the Universe. Considering the case of motion in one dimension say along the $x-axis$, if a particle happens to be at a point $x_{1}$, at time $t_{1}$, and at a later time $t_{2}>t_{1}$, this particle is located at $x_{2}$, we know that the speed $V$ of this particle  is given:

\begin{equation}
V=\left|\frac{\Delta x}{\Delta t}\right|=\left|\frac{x_{2}-x_{1}}{t_{2}-t_{1}}\right|.
\end{equation}

\noindent It is clear from the above that if there exists no limits on the intervals: $\Delta x=x_{2}-x_{1}$ and $\Delta t= t_{2}-t_{1}$, that particle's speed will range from zero to infinity. That is, for any finite duration: $\Delta t>0$, for which: $\Delta x=x_{2}-x_{1}=0$, we will have $V=0$ and for any finite separation: $\Delta x>0$, for which $\Delta t= t_{2}-t_{1}=0$, we will have $V=\infty$, hence: $0\leq V \leq \infty$. So far, so good, no problem -- lets proceed!\\

If we set a minimum time interval, say $t_{p}$, such that for all $t_{2}>t_{1}$, $\Delta t>t_{p}$, where $t_{p}$, is smallest possible interval of time; then, for any space interval: $\Delta x= x_{2}-x_{1}$, there will exist a maximum speed for that particular space interval, let us write this as $V_{max}(x_{2},x_{1})$, and this will be given:

\begin{equation}
V_{max}(x_{2},x_{1})=\frac{\left|x_{2}-x_{1}\right|}{t_{p}}.
\end{equation}

Additionally, if there exists a minimum distance that any two points can ever come closest; that is, the points $x_{2}$, and $x_{1}$, can be brought closer together up until a certain minimum, call this minimum $\ell_{p}$, then, we can talk of an absolute maximum speed, $V_{amax}$, between any two-points of space. This absolute maximum speed, call it $c$, is, unlike $V_{max}(x_{2},x_{1})$,  independent of the coordinates hence thus for any object moving in such a spacetime endowed with space limits:

\begin{equation}
V\leq c=\frac{\ell_{p}}{t_{p}}.\label{v}
\end{equation}

Any abject that travels at this speed $c$ is basically traveling the minimum possible distance in the least possible time duration or it travels an integral multiple ($n\ell_{p}:n=1,2,3 ...$) of this distance in an integral multiple time of the least time ($nt_{p}:n=1,2,3 ...$). From the above thesis, what this means is that spacetime must have space and time limits if it is to have a universal and absolute maximum speed; that is, for any two points $x_{2}$, and $x_{1}$, and any two-points on the $time-axis$ $t_{2}$, and $t_{1}$, the following must hold: $x_{2}-x_{1}\geq \ell_{p}$, and $t_{2}-t_{1}\geq  t_{p}$. From this, the justification of the limits placed on the function $\delta_{\mu}(t)$ [see (\ref{dirich})] follows smoothly.

The above simple reasoning that has lead us to conclude the implied existence of a minimum length and time has a deeper meaning for it tells us that the STR implies a minimum possible time and a minimum possible length! The meaning is deep given the efforts by Giovanni Amelino-Camelia ($2002$) who has proposed a theory -- known as the the Doubly Special Relativity (DSR), that seeks to extend the STR. This theory's hope (should it be confirmed by experiments) is to supersede the STR. The DSR proposes a new observer-independent scale-length. Giovanni Amelino-Camelia proposed this new observer-independent scale-length completely unaware of the fact just presented that the STR implies a minimum length. As discussed below, he [Giovanni Amelino-Camelia]  did this [to propose the existence of  minimum length] to solve a problem to do with the Quantum Gravity (QG) regime.

Basing their arguments on logical intuition and known Laws \textit{of} Physics -- democratically, researchers in the field of QG do agree that at a special scale-length known as the Planck scale-legnth, $\ell_{p}$: a full theory of QG is needed to describe the physics at this scale. However -- according to the STR, different observers  (depending on their state of motion) will measure different lengths, thus they will (may) not agree on whether or not a particle has reached its Planck length. This  presents a ``puzzle/paradox'' for the STR. If they agreed on the Planck scale, then their motions must be similar. If their motions are dissimilar and they agreed on the Planck scale, it would mean the Laws \textit{of} Physics must be different for different observers. This goes against the very foundations of the STR -- clearly, this is unacceptable! To solve this, Giovanni Amelino-Camelia proposed his DSR theory which has been welcome by a significant number of researchers (see e.g. Kowalski-Glikman $2003$; Magueijo \& Smolin $2002a,b$).

Now, given that we have shown that the STR implies a minimum length, this clearly puts the effort of Giovanni Amelino-Camelia into question. Actually, this renders the DSR as an unnecessary effort as it  tries to address something already implied by the STR. 

In closing this section, we need to mention that beginning in the third paragraph from the preceding, we have digressed from the main theme of this section, so -- to keep the reader on track, we shall remind ourself that all we wanted to do has been to show that the existence of an upper cosmic speed limit implies the existence of  a minimum length and a minimum time interval. With this reminder, we proceed to the next section where we hint at the uncertainty principle of QM.

\subsection{\subsectionfont Spacetime Fluctuations/Uncertainty}

An important point to take note of is that, for a point $x_{k}$ on the CST, we are $100\%$ sure where to locate this point, we simple go directly to the point $x_{k}$ and find it there -- this is not true for the QST because $\delta_{k}(t)$ is random and dynamic function -- no one and absolutely no one knows what its value at any given time will be. Given the quantum mechanical uncertainty principle and its intrinsic random nature, could it be that the QST setforth here, could be the answer to randomness and the probabilistic nature of QM? We see here that the point $x_{\mu}$ when cast onto the QST -- $X_{\mu}$, it will be uncertain by a minimum limit of $\ell_{p}$ -- that is to say, this point is cast randomly on the QST. The Laws \textit{of} Nature must be written on the QST and not on the CST. From this, it is clear that there will be an uncertainty ($\Delta x_{\mu}$) in the point $x_{\mu}$,  and this uncertainty:

\begin{equation}
\Delta x_{\mu}\geq\ell_{p}\delta_{\mu}(t).\label{x-delta}
\end{equation}

The deeper meaning of what this means is that any-point on the CST -- $x_{\mu}$, can be cast anywhere on the QST -- $X_{\mu}$, expect inside a small hyper-volume sphere of radius $\mathcal{R}_{\mu}=\ell_{p}$ on the QST. This is very interesting and the meaning of it -- I should admit, is beyond me at present. I would like to set this as an area for further research.

\subsection{\subsectionfont  Energy-Momentum Fluctuations/Uncertainty}

Taking the time derivative of the coordinates of the QST, we have: $\dot{X}_{\mu}(x)=\dot{x}_{\mu}+\ell_{p}\dot{\delta}_{\mu}(x)$, and this implies: $\mathcal{P}_{\mu}=p_{\mu}+m\ell_{p}\dot{\delta}_{\mu}(x)$ where $m$ is the mass of the particle whose momentum is $\mathcal{P}_{\mu}$; and comparing this with: $\mathcal{P}_{\mu}=p_{\mu}+\Lambda_{\mu}\hbar$, this means: $m\ell_{p}\dot{\delta}_{\mu}(x)=\Lambda_{\mu}\hbar$, hence thus the $4$-vector cosmological field can be written:

\begin{equation}
\Lambda_{\mu}=\left(\frac{m\ell_{p}}{\hbar}\right)\dot{\delta}_{\mu}(t),
\end{equation}

hence thus the field: $\Lambda_{\mu}=\Lambda_{\mu}(t)$, will have the properties of the field $\dot{\delta}_{\mu}(t)$, the meaning of which is that $\Lambda_{\mu}$, will be a random-dynamic field since $\dot{\delta}_{\mu}(t)$, is a random and dynamic field. This means the fluctuation of the energy of the particle will be $\Delta\mathcal{E}_{\mu}=\Lambda_{\mu}\hbar$. We are not going to determine here the limits of the fluctuation in the energy as this will require us to go deeper into the nature of the QST -- we shall reverse this for a separate reading. 

\section{\sectionfont C-Violation: Possible Reason for Matter-Antimatter Asymmetry\label{ctv_s}}

In the Dirac Theory, antiparticles are negative energy particles which travel back in time and these have the opposite electronic charge of their particle counterpart. In this theory [Dirac], the Dirac Equation is symmetric under electronic charge conjugation. What this means is that the same law, or more clearly, the same equation that governs particles governs antiparticles. In the present theory, we see that $\rm{C}$-\textit{symmetry} is violated, the meaning of which is that particles are not governed by the same equation that governs particles and vis-versa.

In this case where particles and antiparticles are governed by different laws, if these laws are defined on the same spacetime continuum, then, nothing stops particles and antiparticles from existing in the same spacetime continuum. In this kind of set-up, we would expect to observe particles and antiparticles in equal proportions in the same region of space at all times. If -- as in the present set-up, these laws are defined on separate spacetime continuums (one in which $\Lambda_{0}>0$ and the other in which $\Lambda_{0}<0$: we have advanced this in \S \ref{at}), then, matter and antimatter will exist in separate regions of space at altimes. Simple -- there will exist an apparent asymmetry in the distribution of matter and antimatter.

It must be clear that for any spacetime continuum, the corresponding cosmological field along a given axis  can  only either be positive or negative and never both -- remember the cosmological field can either be positive or negative along each of the axis -- see equation (\ref{cosm_const}). For example, in the case of the time component of the cosmological field, either $\Lambda_{0}>0$ or $\Lambda_{0}<0$ and never can we have both fields in the same region of the spacetime continuum -- it is just plain meaningless because they could cancel each other rendering its inclusion purposeless. So, what this means is that we are going to have two physically separate spacetime continuums, one in which $\Lambda_{0}>0$ and the other in which, $\Lambda_{0}<0$. 

\begin{quote}
\textbf{\textsl{Hence thus, it must be clear from this that the inclusion of the cosmological field in the time dimension leads to an explanation of why matter and antimatter will not be seen to exist in the same region of spacetime.}}
\end{quote}

\begin{table}[h!]
\caption{\tabletitlefont Signature of  the New Spacetime\\}
\label{tspace}
\begin{tabular}{|c |c c c | c c c| c c |}
\hline
$\Lambda_{k}\longrightarrow$ & $\omega_{1}$ & $\omega_{2}$ & $\omega_{3}$ & \multicolumn{3}{c|}{\textbf{Space Coordinates}} & \multicolumn{2}{c|}{\textbf{\underline{Particle Energy}}} \\
  & & & & \multicolumn{3}{c|}{\textbf{\underline{}}} & $\mathcal{E}>0$ & $\mathcal{E}<0$\\
\hline\hline\hline
\textbf{Region $1$} & $+1$ & $+1$ & $+1$  & ($x>0$, & $y>0$, & $z>0$) & ($\mathcal{E}_{1},|\mathcal{E}_{17}|$) & ($-\mathcal{E}_{1},\mathcal{E}_{17}$)\\
\textbf{Region $2$} & $-1$ & $+1$ & $+1$  & ($x<0$, & $y>0$, & $z>0$) & ($\mathcal{E}_{2},|\mathcal{E}_{16}|$) & ($-\mathcal{E}_{2},\mathcal{E}_{16}$)\\
\textbf{Region $3$} & $+1$ & $-1$ & $+1$  & ($x>0$, & $y<0$, & $z>0$) & ($\mathcal{E}_{3},|\mathcal{E}_{15}|$) & ($-\mathcal{E}_{3},\mathcal{E}_{15}$)\\
\textbf{Region $4$} & $-1$ & $+1$ & $-1$  & ($x>0$, & $y>0$, & $z<0$) & ($\mathcal{E}_{4},|\mathcal{E}_{14}|$) & ($-\mathcal{E}_{4},\mathcal{E}_{14}$)\\
\textbf{Region $5$} & $-1$ & $-1$ & $+1$  & ($x<0$, & $y<0$, & $z>0$) & ($\mathcal{E}_{5},|\mathcal{E}_{13}|$) & ($-\mathcal{E}_{5},\mathcal{E}_{13}$)\\
\textbf{Region $6$} & $-1$ & $+1$ & $-1$  & ($x<0$, & $y>0$, & $z<0$) & ($\mathcal{E}_{6},|\mathcal{E}_{12}|$) & ($-\mathcal{E}_{6},\mathcal{E}_{12}$)\\
\textbf{Region $7$} & $+1$ & $-1$ & $-1$  & ($x>0$, & $y<0$, & $z<0$) & ($\mathcal{E}_{7},|\mathcal{E}_{11}|$) & ($-\mathcal{E}_{7},\mathcal{E}_{11}$)\\
\textbf{Region $8$} & $-1$ & $-1$ & $-1$  & ($x<0$, & $y<0$, & $z<0$) & ($\mathcal{E}_{8},|\mathcal{E}_{10}|$) & ($-\mathcal{E}_{8},\mathcal{E}_{10}$)\\
\hline
\end{tabular}
\end{table}

Further, this fact that the corresponding cosmological field along a given axis  can  only either be positive or negative and never both implies that each of these spacetime continuums shall have to be sub-divided into $8$ different sections since we have $3$ axis and along each of these we have two choices for $\Lambda_{k}$ -- that is, either $\Lambda_{k}>0$ or $\Lambda_{k}<0$. The the laws of permutations and combination dictate that we must have $8$ different combinations and this is shown in table (\ref{tspace}). In this table, column $1$ gives the section or sector of space and columns $2,3,4$ gives the corresponding sign of the cosmological field along the $x,y,$ and $z-axis$ and columns $5,6,7$ list the segment on which a given point falls on this space continuum and lastly, columns $9,10$ give the energy of the particles that will be found in that sector of space. We shall explain in \S (\ref{t_s}), why these particles fall in that space segment we have placed them.


\begin{figure}[ht!]
\begin{center}
{\tabletitlefont Eight Segment Sectioning of the $3\rm{D}$ Space}
\shadowbox{\epsfysize=6.0cm \epsfbox{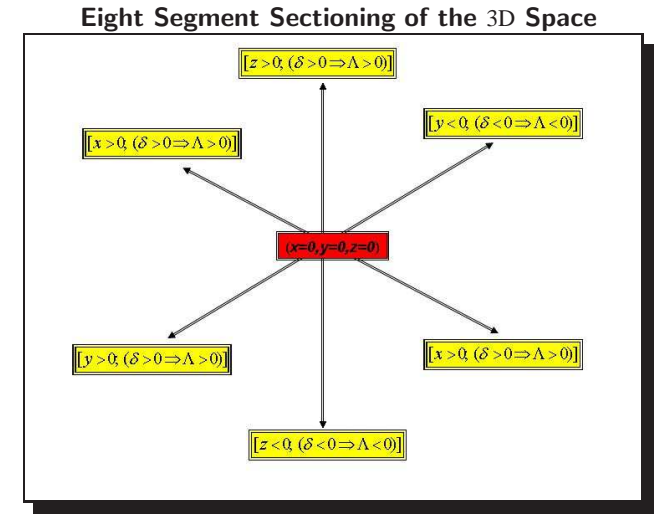}}
\end{center}
\caption{\figtextfont A schematic diagram illustrating the $8$-segment sectioning of the of the $3\rm{D}$ space. For the case $\Lambda_{0}>0$, this entire $3\rm{D}$-space will be filled at each and every-point with the  cosmological field $\Lambda_{0}>0$, and likewise for the case $\Lambda_{0}<0$, this entire $3\rm{D}$-space will be filled at each and every-point with the  cosmological field $\Lambda_{0}<0$.}
\label{ds}
\end{figure} 

The ``picture'' of the structure of emergent space is shown in figure (\ref{ds}). This picture illustrates the $8$-segment sectioning of the of the $3\rm{D}$ space. Each axis is endowed with the corresponding field $\delta_{k}$ and we have choose the sign of this field to be the same as the sign on the corresponding coordinate. If, we consider here only positive energy particles, then, in each of the regions, we would expect two particles of different masses, much like what we see with the Electron and the Muon. The masses of these particles for the different regions will be different. We will not go any further in this subject of the particles to be found in the different regions of the $8$-segment space but leave this for a fresh reading. All we hope is that the reader does see the potency and the hidden veracity to be found in the ideas propagated herein.

\section{\sectionfont Darkmatter and Darkenergy}

The subject of \textit{Darkmatter} and \textit{Darkenergy} is a hot topic of intense theoretical and experimental research (see e.g. Sofua $1997$; Sofua \textit{et al.} $1997$; Sofua \textit{et al.} $1997$). We find that our search -- for a solution to the problem of why the Universe appears to be composed of matter with no significant quantities of antimatter, has unexpectedly, lead us also to the problem of darkmatter and darkenergy. The presence of darkmatter/darkenergy first come to light  in 1933 when the Swiss  astronomer Fritz Zwicky  conducted observations of galaxies in the Coma Cluster -- the massive cluster of galaxies closest to the Milkway Galaxy -- where he was able to convincingly demonstrate that, given their masses, these galaxies where moving unexpectedly fast relative to one another -- so much that, they should have escaped their gravitational influence - but, for some strange and unknown reason,  they had not done so. This presented a puzzle because Newtonian gravitation theory does not agree with this.

There are two possible solutions to this puzzle: either the stars that we observe are merely a tracer (about $1\%$) of the total amount of matter in the cluster and the rest is in the form of some kind of exotic ``dark'' matter; or gravity is much stronger on million light-year scales than the expected Newtonian $r^{-2}$ force law. However, neither Zwicky's observations nor those made in the intervening years have allowed researchers to distinguish conclusively between these solutions. Actually, the presence of darkmatter/darkenergy is so wide spread through the cosmos so much that something is missing in our understanding of the Laws \textit{of} Nature. 

As to the question of what this darkmatter/darkenergy is, there exists a plethora of ideas. Our ideas join in the ranks and file of these ideas -- all in the effort to finding answers to this great cosmic mystery. It should be said here that we find no reason to discuss the other ideas that have been proposed because we believe the present ideas stand on their own.

We have already suggested that the vacuum can be assigned an energy $\mathcal{E}_{vac}=\Lambda_{0}\hbar c$. Within the framework of equation (\ref{fsol}), this energy assignment corresponds to the momentum solution:

\begin{equation}
\mathcal{P}^{k}\mathcal{P}_{k}c^{2}+\epsilon ^{2}Q^{2}c^{2}\equiv0,\label{tach1}
\end{equation}

and -- further, this corresponds to a wavefunction ($\psi_{D}$) for a particle that satisfies  the following wave equation with decoupled energy and momentum fields:

\begin{equation}
\begin{array}{c  c c c}
i\hbar\gamma^{0}\partial_{0}\psi_{D}+i\hbar \gamma^{0}\Lambda_{0}\psi_{D}=0 & ...
& (\textbf{a})\\
\\
i\hbar\gamma^{k}\partial_{k}\psi_{D}+i\hbar\gamma^{k}\Lambda_{k}\psi_{D}=\epsilon Q\psi_{D} & ... & (\textbf{b})
\\
\end{array}.\label{tach2}
\end{equation}

It is not difficult for one to see that equation (\ref{tach1}) implies that the particle's momentum will be imaginary! This momentum is given: $p_{k}=-\omega_{k} \Lambda_{k}\hbar \pm i \epsilon Qc/\sqrt{3}$. We know from the STR that a particle with imaginary momentum will have to travel at speeds greater than the speed of light. These particles, that travel faster than the speed of light are known as Tachyons and are at present nothing but hypothetical particles born out of a deep theoretical curiocity; they [Tachyons] have never been directly or indirectly observed. Discarding Tachyons as mere unphysical means we have to consider neutral particles, i.e. $Q=0$ in which case the particle $\psi_{D}$ will have real momentum. Thus this particle -- $\psi_{D}$, moves at the speed of light and has no electronic charge (rest-mass). We would like to think of this particle -- $\psi_{D}$, as a darkparticle.

The reason for suggesting that the particle $\psi_{D}$, be a darkparticle is a simple one. This particle's four momentum, which is given by $p_{\mu}=\Lambda_{\mu}\hbar$, coinsides with  $4$-vector cosmological field that we have just added. We added this $4$-vector cosmological field with the hope that it will be a property of the vacuum. Now, we realize that this $4$-vector cosmological field describes a particle. This particle -- as the vacuum, must be all-pervading and permeating, thus it must be elusive and its physical presence most certainly will be observed via the effects of its energy and momentum field. The $4$-vector cosmological field is a well behaved random field and this property (of randomness) fits well in describing random fluctuations of the vacuum. This descriptions, leads one to the idea that this particle must be a darkparticle. 

\textbf{\underline{Proposal}:} \textsl{This description above of $\psi_{D}$ -- in our view or opinion,  suits the description of a darkparticle -- hence thus, we propose that the dark-energy-momentum that is thought to fill all of space, is comprised of this particle.} 

The inclusion of darkmatter and darkenergy has implications on the way gravitation works. In a seperate reading whose work is currently underway, we shall address this problem. For now, we hope the reader will be content with what we have presented.

\section{\sectionfont A New Model of the Vacuum\label{vac_m}}

In 1930, two years after he proposed his relativistic wave equation, Dirac had to face head-on the inevitable fate of the Dirac Electron foretold by his theory. The intrinsic and inherent spacetime symmetries embodied in the usual mundane CST on which the STR is built that extend to Dirac's Theory, meant negative energies would exist and these extended downwards with no limits in a sort of mirror image of the positive energy levels. As aforementioned -- in the formulation of his theory, these negative energies, is what Dirac had hoped to deracinate. He thought, the negative probabilities exhibited by the Klein-Gordon theory is what lead to the negative energies, thus he reasoned that eliminating these, would in one full-swap, eliminate the negative energies -- he was wrong! Much to his own chagrin and that of others, the negative energies reared their head in the new Theory \textit{of} Dirac, thus bewildering any effort to reed ourself of them.

As is now bona-fide knowledge, what the negative energies of the Dirac Electron really meant is that the usual ground state of say the Hydrogen atom, is not really the true ground state at all but has a bottomless pit of negative states. If this where the case, the Electron would have to fall eon to the bottom of the bottomless pit and in the process endlessly emit energy. This could mean the Electron in the Hydrogen atom must be a source of new matter and energy as it could create more and more energy as it journeys forever down to the bottom of the baseless pit. Seen from the other side of the veil, this in actual fact meant that matter should be inherently unstable. By any stitch of imagination, this is a fact not supported by observations thus Dirac had no choice but to face this problem head-on. To solve this problem, Dirac had to redefine the vacuum otherwise nothing of the beautiful equation he had discovered could remain as it could have been found in serious contempt of physical and natural reality!

\begin{widetext}

\begin{figure}[h]
\begin{center}
\shadowbox{
$\begin{array}{c c}
\multicolumn{1}{l}{\mbox{}} &
\multicolumn{1}{l}{\mbox{}} \\ [-0.53cm]
\epsfysize=7.0cm
\epsfbox{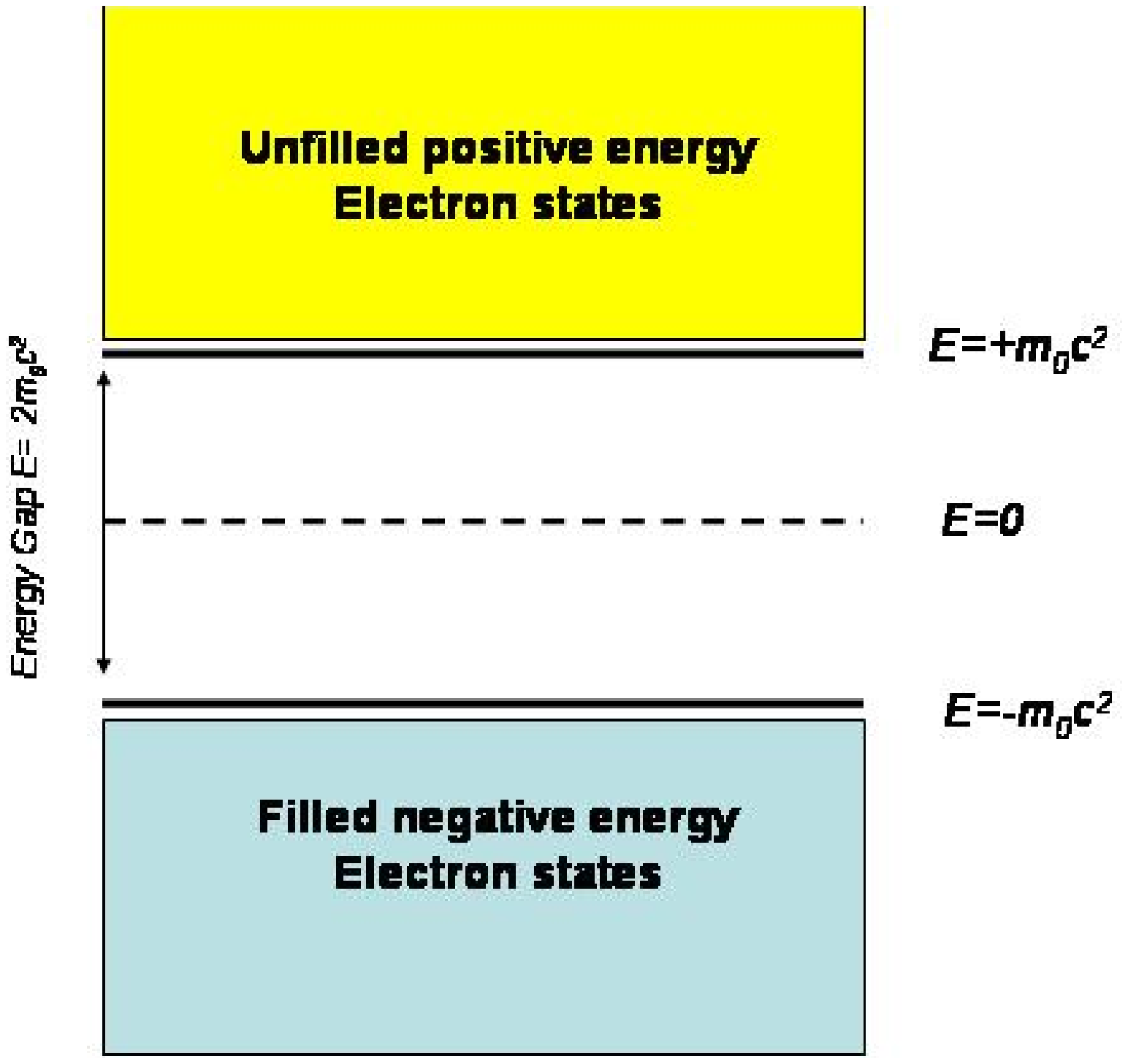} &
\epsfysize=7.0cm
\epsfbox{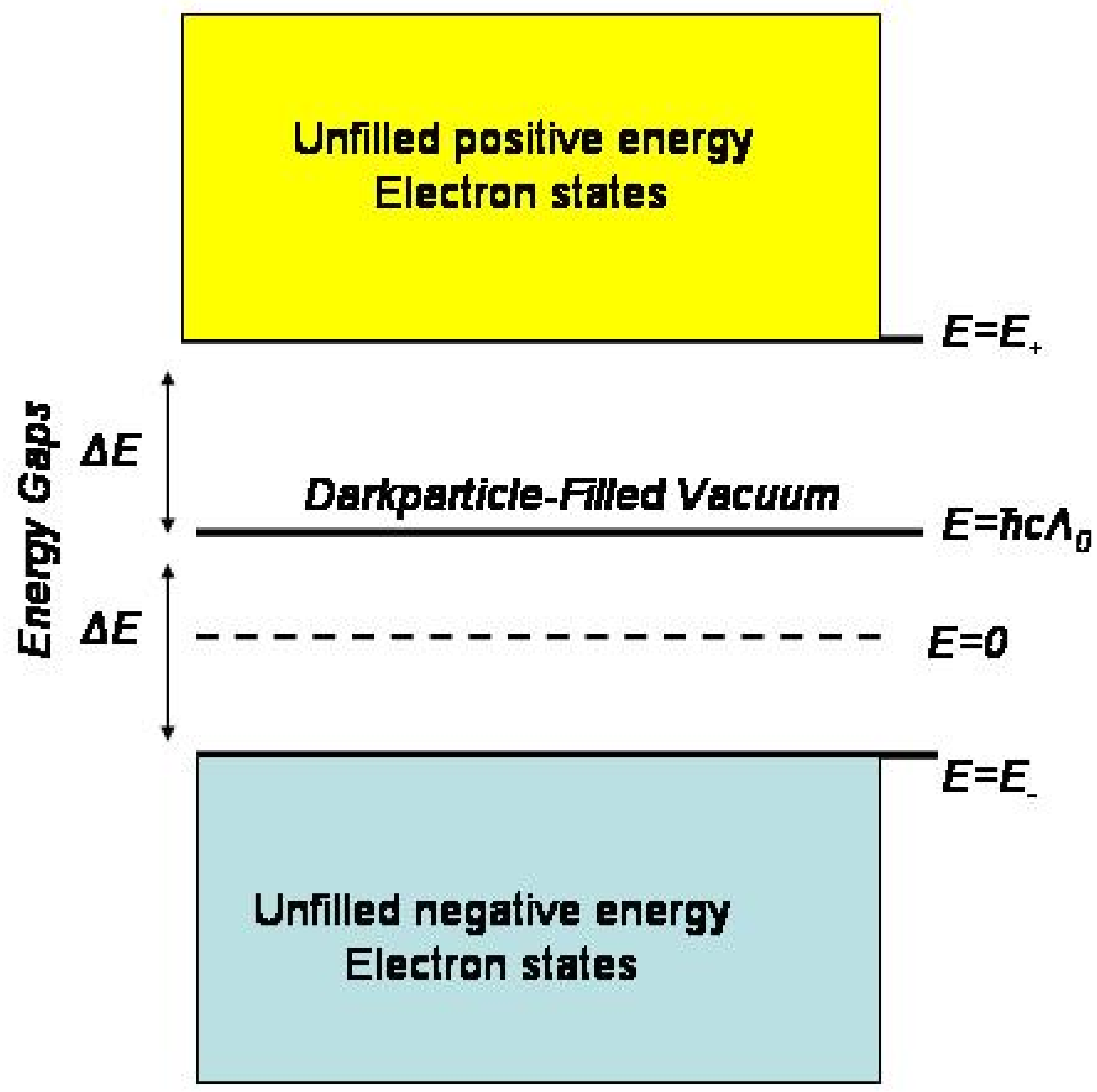} \\ [0.4cm]
\mbox{\bf (a)} & \mbox{\bf (b)} 
\end{array}$}
\end{center}
\caption{\figtextfont (\textbf{a}) Dirac's model of the vacuum with a sea of negative energy electrons occupying all the negative energy states and (\textbf{b}) The proposed new model of the vacuum with unfilled negative energy states and having a finite energy composed of pairs of tachyons. These pairs of tachyons have opposite electronic charges thus the vacuum has net zero electronic charge and at the same time, its energy is finite. These tachyon are in a constant state of annihilation and creation. }
\label{vacm}
\end{figure}
\end{widetext}

At the time Dirac had to redefine the vacuum, it was thought and taken as a self-evident, most logical and rational truth beyond question that the vacuum contained complete nothingness -- thus, this effort by Dirac to give the vacuum physical properties was nothing short of a revolution in thinking. He defined the vacuum to consist of unfilled positive and filled negative energy states of the Electron. According to the Pauli exclusion principle, an Electron would be prevented from making a downward transition if all the negative energy states are occupied hence thus, the positive energy Electron where spared their dooms fate -- that of falling into the endless pit of negative energies. The ``picture'' of this vacuum model is shown in figure (\ref{vacm}) (a).

The only problem with the Dirac vacuum (also known as the Dirac sea) is that it must have an infinite negative energy and infinite electronic charge! As to why we do not ``see'' the infinite electronic charge and energy of this vacuum, Dirac proposed that this would have to be invisible and beyond the realm of measurement. Although this model of the vacuum has great predictive powers, in that the existence of antiparticle is implied and also some of the predictions of QED that have given it widespread acceptance as one of the best theories we have, it has some problems. Actually, the vacuum of QED is different from that of Dirac but retains some features of the Dirac vacuum. 

The vacuum of QED contains what is known as the zero-point energy, which is a finite  intrinsic minimum none-zero energy contained in the vacuum. This energy continuously transforms some of its energy into mass causing the emergence of random appearance and disappearance of electronically charged particle(s)-antiparticle(s) pairs and these are known as virtual particles. The virtual particles can not be observed in real life just as the Dirac vacuum.

We do not object but say -- \textit{yes}, the Dirac vacuum, has had much success, \textit{but} despite this -- true is that, the idea of the Dirac vacuum [\textit{viz}, its \textbf{\textit{infinite}} electronic charge and energy together with the fact that this (charge and energy) must be unmeasurable and invisible] tends not to strike the layman as well as the esoteric, as very elegant. Another problem is; why should the negative energy Electron be invisible and unmeasurable? As long as this electronic charge is unmeasurable, it is permanently safe as an idea since it can not be refuted! This serious desiderata goes against the true and noble spirit of science, since science concerns its self with physical phenomena that can be measured and falsified and anything beyond the realm of measurement and falsification is beyond the realm of science as-well, it is something else not science. 

I will say nothing further, \textit{viz} the Dirac vacuum, lest I may appear to be discrediting this vacuum in the hope of replacing it with a new one -- no! I am simple stating the natural mystification one feels derived from this fact of unmeasurable and invisible Dirac charge and energy. I shall leave the Dirac vacuum here and proceed to give the emergent vacuum model from the present theory.

In complete harmony and resonance with the idea of the QED vacuum model where we have a zero-point energy, the modification we have made to the Dirac Equation enables us to assign a vacuum energy, that is: $\mathcal{E}_{vac}=\mathcal{E}_{9}=\Lambda_{0}\hbar c$.  Already pointed in the closing part of the last section, the particle with this energy solution is a darkparticle. Unlike the QED vacuum, these darkparticles are not virtual but real. These darkparticles act a seal to prevent positive energy particles from falling into the negative energy state. To see this, suppose we have an Electron in the energy state: $\mathcal{E}_{j}>\mathcal{E}_{vac}: j=1,2, ..., 8$ and this Electron is to make a downward transition to a negative energy state, it would have first to make a transition to the vacuum energy state $\mathcal{E}_{vac}$. Now here is the catch. For any particle of finite rest-mass to make the transition: $\mathcal{E}_{j}\longrightarrow\mathcal{E}_{vac}:j=1,2, ..., 8$, it would have to travel at the speed of light. As we already know from the STR, a material particle of finite rest-mass, first moving at sub-luminal speeds is forbidden from passing the light-speed barrier because this would require an infinite amount of energy! Thus the transition: $\mathcal{E}_{j}\longrightarrow\mathcal{E}_{vac}:j=1,2, ..., 8$, is impossible hence thus, particles will be forbidden by the light-speed barrier from entering the negative energy states! 

A second and much stronger reason is that, if say by shear chance, the Electron manages to get this infinite energy and reaches the light-speed, it would have -- against the Law \textit{of} Conservation of electronic charge; to lose its electronic charge because the vacuum state is only occupied by particles whose electronic charge is zero (or zero rest-mass). Accepting the above thesis, means within the framework of the present ideas, we have solved the problem of why positive energies will be forbidden from making a transition to negative energy states and also how the negative energy states of the vacuum can stay permanently empty without any problem whatsoever. 

The light-speed barrier, $v<c$, and the conservation of electronic change save the day, as the dilemma faced by Dirac is not anymore present -- thanks to these two Physical Laws. We can now have empty negative energy states and these can not be occupied because having just one negative energy state occupied will mean this Electron will have to suffer the fate as the Dirac Electron and fall eon to the bottom of the bottomless pitch of the negative energy well and inthe process creating energy endlessly. With the negative energy states empty and the vacuum filled by the darkparticles -- in my modest opinion, we have the perfect vacuum! The picture of this vacuum model is shown in figure (\ref{vacm}) (b) and this [vacuum] needs no infinite energy and it needs no infinite electronic charge but simple has the light speed barrier and the conservation of electronic charge to take care of the troubles that bedeviled Dirac.

\section{\sectionfont T-Symmetry \label{t_s}}

The fact that the Full Cosmological Dirac Equation (\ref{fsol}) is $\rm{T}$-symmetric, means that we can flip the positive and negative energy solutions about the vacuum energy: $\mathcal{E}_{vac}=\Lambda_{0}\hbar c$, that is to say, the energy solutions: $\mathcal{E}_{1}, \mathcal{E}_{2}, \mathcal{E}_{3} ...., \mathcal{E}_{9}, \mathcal{E}_{10},\mathcal{E}_{11},\mathcal{E}_{12}, ..., \mathcal{E}_{17}$, lead after: $t\longmapsto -t$ or $\mathcal{E}\longmapsto-\mathcal{E}$, to the energy solutions: $|\mathcal{E}_{17}|, |\mathcal{E}_{16}|, |\mathcal{E}_{15}| ...., \mathcal{E}_{9}, \mathcal{E}_{8},-\mathcal{E}_{7},-\mathcal{E}_{6}, ..., -\mathcal{E}_{1}$ where the operator $|[]|$ is the usual absolute operator which gives the absolute value of the quantity inside the operator $[]$. From a symmetry view-point, the latter set of energy solutions is obtained after a flipping of these energies about the vacuum energy. 

This means doublets emerging from equation (\ref{fsol}) will have energies: ($\mathcal{E}_{1},|\mathcal{E}_{17}|$),  ($\mathcal{E}_{2},|\mathcal{E}_{16}|$), ($\mathcal{E}_{3},|\mathcal{E}_{15}|$), ($\mathcal{E}_{4},|\mathcal{E}_{14}|$), ($\mathcal{E}_{5},|\mathcal{E}_{13}|$), ($\mathcal{E}_{6},|\mathcal{E}_{12}|$), ($\mathcal{E}_{7},|\mathcal{E}_{11}|$) and ($\mathcal{E}_{8},|\mathcal{E}_{10}|$), these can be generalized as ($\mathcal{E}_{j},|\mathcal{E}_{18-j}|$) and for the negative energies, we will have ($-\mathcal{E}_{j},\mathcal{E}_{18-j}$). As has been be argued in \S (\ref{vac_m}) above, each of the different Universe, $\mathcal{U}^{+}$ and $\mathcal{U}^{-}$ are filled exclusively with positive and negative energy particles respectively. These particles will belong to the different section of the two Universes $\mathcal{U}^{+}$ and $\mathcal{U}^{-}$ and this is shown in the forth block of table (\ref{tspace}). 

In each of the regions of spacetime, a particle must have its other doublet pair which is not the same mass as its self but has all other properties the same. This brings to mind the Electron and Muon which appear similar in all aspects except their mass. We simple want to point this out that the present theory contains such interesting information and that the time to delve into a full throttle on this is not now as this is better done in further readings that expand on the present.

\section{\sectionfont Discussion and Conclusions}

If the reader has gone through this reading up-until the present point, I sincerely believe they [the reader] will agree with me if I say that the intent of this reading -- to address the matter/antimatter asymmetry has been dwarfed, or pretty much appears much less significant when compared to what we have actually discovered along the way. The initial intent was a rather modest ambition -- to use the inclusion of the cosmological field in the time operator to  explain the apparent asymmetry as to why the Universe seems to be predominately composed of matter instead of equal portions of matter and antimatter as predicted by the Dirac Equation. Along the way on the voyage, we found that; (1) we had to propose a new model of spacetime that has the potency to explain the quantum mechanical uncertainty and randomness; (2) we had to setforth a new model of the vacuum radically different from that of Dirac; (3) the modified Dirac Equation allows for the existence of a spectrum of eight particles with unique masses and these reside in pair in the 8-different segments of space as layed-out in section \S (\ref{ctv_s}); (4) the predicted vacuum by the present theory is that composed of all-pervading and permeating particles and these particles may explain why they appear to be darkmatter and darkenergy in the Universe.

Further, allow me to say that, given the aforesaid and if correct, and the present theory is anything to go by, that it corresponds to natural reality, then, the modification made here to the Dirac Equation is so simple and trivial yet very deep. I should say, in my perusal through the literature that I have so far been able to lay my hands, I have not come across a modification of this kind. Concurrently, I have not come across an approach where the inclusion of a cosmological $4$-vector field in the spacetime operator (as has been done here) is used to probe space and time asymmetries/symmetries. It may so happen that someone already has done this kind of work given its simplicity and trivial nature. If this is the case, I wonder why for example it is considered a big mystery that the Universe appears to be made up chiefly of matter and that the best explanation we have of this are the ideas laid down by Andrei Sakharov in $1967$ where Nature must adhere to a strict prescription inorder to explain this mystery. 

One may ask: If Andrei Sakharov had in $1967$ set conditions to explain the apparent matter/antimatter asymmetry -- why the present work to champion the same endeavor? We feel that these conditions are far too many (four) and that if a simpler solution can be found, then it must be found and set into motion as a contender to prevailing wisdom. On another level , there is a general feeling amongst a good number of researchers that the currently measured $\rm{CP}$\textit{-violation}s fall far too short to explain this apparent matter/antimatter asymmetry (see e.g. Rodger $2001$; Sinha $2009$). Because of the said reasons and others not mentioned here, I felt it was time we seek alternative ideas -- hence the present.

Clearly, the bedrock or the very foundations of our new theory rest entirely on the modification we have made to the CST which has been to transform it into a new spacetime endowed with a $4$-vector cosmological field and we have called this new spacetime -- the QST. This 4-vector cosmological field has the property of  pure randomness (unpredictability). In this QST, it is not possible to know exactly  where a particular point from the CST will be located upon a transformation to the the QST as this attribute is random -- thanks to the function $\delta_{\mu}(t)$. The QST clearly defies the \textit{sacrosanct} $\rm{CPT}$ theorem of L\"uder, Bell and Pauli, that every Lorentz invariant theory must observe $\rm{CPT}$-\textit{symmetry}. Clearly and without any doubt, the derived Cosmological Dirac equation is Lorentz invariant yet it violates this  sacrosanct $\rm{CPT}$ theorem. Not only is this equation in violation of $\rm{CPT}$, it is in violation of all the symmetries with the exception of the $\rm{T}$-\textit{symmetry}.

The new vacuum model setforth in this reading is radically different from that setforth by Dirac in that a particle's negative energy levels -- just as the positive energy levels, need not be filled. The vacuum has an energy level $\mathcal{E}=\Lambda_{0}\hbar c$ (constituting of electrically neutral darkparticles that travel at the speed of light) and this energy level acts as a barrier that can prevents positive energy particles from falling into the negative energy well and vice-versa. If a positive energy particle where to make a transition to a negative energy state, it will need to first enter in this darkparticle meaning to say it must at somepoint travel at the speed of light. From the STR, we know that particles traveling at sub-luminal speed will need an infinite amount of energy to be propelled to the light-speed -- the meaning of which is that it must be impossible for a particle to be propelled to light-speeds hence thus, in this way, the light speed barrier plays an very important role in the stabilizing the vacuum. 

We also saw that these darkparticles can actually carry some electrical charge but for this to be so, they will have to travel at superluminal speed. It may well be possible that these charged darkparticles may be real. For no other reason except that we wanted to keep the matters as simple as one can, we choose to discard these and this may not be the correct thing to do.

Furthermore, we saw that excluding the vacuum particles, the modified Dirac Equation predicts a total of $16$ energy levels which invariable means $16$ particles with $8$ of these energy levels being positive while the other energy levels are negative. But given that the vacuum setforth here separates the positive and negative energy particles, one can only talk of $8$ particles and these occupy the eight different region of space as. If one combined the present ideas with the three curved spacetime equation derived in the reading Nyambuya ($2008a$) -- in my view, this positions us on pedestal to perharps explain why fundamental particles have varying masses and why they the generation of leptons come in pairs. 

Insofar as symmetry is concerned, we have in the present reading, destroyed the beautiful Dirac-Symmetries and amongst these and top of the list, is the cornerstone and sacrosanct $\rm{CPT}$-\textit{symmetry}. Dirac-Symmetries have inspired physicists to seek highly symmetric theories and amongst these is the Supersymmetry Theory known as SUSY whose endeavor is to unite QM with the GTR. In light of what we have presented here, if it is correct, then, clearly highly symmetric theories may not be the desired thing for a final unified theory. This is just my opinion and obviously this may change as more understanding of reality and nature comes to light.

\textit{\textbf{In closing}}, allow me to say that, writing this reading and my other readings Nyambuya ($2007,\, 2008a,b$) has been such pain because in all these readings, I find myself having to introduce new ideas, terms and concepts and this by any measure is no easy task -- because of this, I ask the reader to bear with me.

{\tabletextfont  \underline{\tabletextfont  \textbf{Acknowledgments:}} This work was completed under the kind hospitality of my brother -- George, and his wife -- Sarmatha.  I am grateful for this and as-well to Donald Ngobeni for his support during the drafting of this manuscript. I dedicate this reading to the Pine o\textit{f} Lilyrose.}

\end{document}